\documentclass[fleqn,usenatbib]{mnras}

\usepackage{newtxtext,newtxmath}
\usepackage[T1]{fontenc}

\DeclareRobustCommand{\VAN}[3]{#2}
\let\VANthebibliography\thebibliography
\def\thebibliography{\DeclareRobustCommand{\VAN}[3]{##3}\VANthebibliography}

\usepackage{graphicx}	% Including figure files
\usepackage{amsmath}	% Advanced maths commands

\usepackage{natbib}%,twoopt}
\usepackage{mathtools}

\DeclarePairedDelimiter\abs{\lvert}{\rvert}%
\newcommand{\logrhk}{$\rm log\,R^{\prime}_\mathrm{HK}$}
\newcommand{\kms}{\,km\,s$^{-1}$} % kilometres per second
\newcommand{\mstar}{M$_{\star}$}
\newcommand{\rstar}{R$_{\star}$}

\newcommand{\msun}{$M_{\odot}$}

\newcommand{\vsini}{$v$\,sin\,$i_\star$}

\newcommand{\teff}{$T_{\rm eff}$}
\newcommand{\logg}{log\,{\it g$_\star$}}
\newcommand{\feh}{[Fe/H]}

\newcommand{\mearth}{$M_{\oplus}$}

\newcommand{\mjup}{$M_{\rm J}$}
\newcommand{\smw}{S$_{\text{MW}}$}
\title[The GAPS Programme at TNG LXVIII]{The GAPS Programme at TNG LXVIII. \\ Characterization of the outer substellar companion around HD\,72659 with a multitechnique approach
   \thanks{Based on: observations made with the Italian Telescopio Nazionale \textit{Galileo} (TNG), operated on the island of La Palma by the INAF - Fundaci\'on Galileo Galilei at the Roque de Los Muchachos Observatory of the Instituto de Astrof\'isica de Canarias (IAC). 
   Based on observations collected at the European Southern Observatory under ESO programme(s) 096.C-0241(C) and 112.25JY.001 (SPHERE). 
   Based on observations collected at the European Southern Observatory under ESO programme(s) 183.C-0972, 106.215E, 105.20AK, 099.C-0458, 093.C-0409, 090.C-0421, 085.C-0019, 072.C-0488, 0103.C-0432, 0102.C-0584, 0102.C-0558, 0101.C-0379, 0100.C-0097 (HARPS).}}

\author[A. Ruggieri et al.]{
A. Ruggieri$^{1,2}$\thanks{E-mail: alessandro.ruggieri@inaf.it},
A. Sozzetti$^{3}$,
S. Desidera$^{2}$,
D. Mesa$^{2}$,
R. Gratton$^{2}$,
F. Marzari$^{2,4}$,
M. Bonavita$^{5}$,
K. Biazzo$^{6}$,
\newauthor
V. D'Orazi$^{2,7}$,
C. Ginski$^{8,9,10}$,
M. Meyer$^{11}$,
L. Malavolta$^{2,12}$,
M. Pinamonti$^{3}$,
D. Barbato$^{2}$,
C. Lazzoni$^{2}$,
\newauthor
A. F. Lanza$^{1}$,
L. Mancini$^{3}$,
L. Naponiello$^{3}$,
D. Nardiello$^{2,12}$,
T. Zingales$^{2,12}$,
M. Rainer$^{13}$,
G. Scandariato$^{1}$,
\newauthor
P. Giacobbe$^{3}$,
R. Cosentino$^{14}$,
A. Fiorenzano$^{14}$,
R. Claudi$^{2,15}$
\\
$^{1}$INAF, Astrophysical Observatory of Catania, Via S. Sofia 78, I - 95123 Catania, Italy\\
$^{2}$INAF, Astronomical Observatory of Padua, Vicolo dell'Osservatorio 5, I - 35122, Padua, Italy\\
$^{3}$INAF, Astrophysical Observatory of Turin, Via Osservatorio 20, I - 10025, Pino Torinese (TO), Italy\\
$^{4}$Department of Physics and Astronomy, University of Padua, via Marzolo 8, I - 35131, Padua, Italy\\
$^{5}$Institute for Astronomy, University of Edinburgh Royal Observatory, Blackford Hill, Edinburgh EH9 3HJ, UK\\
$^{6}$INAF, Astronomical Observatory of Rome, Via Frascati 33, I - 00178, Monte Porzio Catone (RM), Italy\\
$^{7}$Department of Physics, University of Rome Tor Vergata, Via della Ricerca Scientifica 1, 00133, Rome, Italy \\
$^{8}$School of Natural Sciences, Center for Astronomy, University of Galway, Galway H91 CF50, Ireland \\
$^{9}$Leiden Observatory, Leiden University, PO Box 9513, 2300 RA Leiden, The Netherlands \\
$^{10}$Anton Pannekoek Institute for Astronomy, University of Amsterdam, Science Park 904, 1098 XH Amsterdam, The Netherlands \\
$^{11}$Department of Astronomy, University of Michigan, 1085 S. University, Ann Arbor, MI 48109, USA \\
$^{12}$Department of Physics and Astronomy "Galileo Galilei", University of Padova, Vicolo dell'Osservatorio 3, I - 35122, Padova, Italy\\
$^{13}$INAF – Osservatorio Astronomico di Brera, Via E. Bianchi 46, 23807 Merate (LC), Italy\\
$^{14}$Fundacion Galileo Galilei – INAF, Rambla J.A. Fernandez P., 7, E-38712
S.C.Tenerife, Spain\\
$^{15}$Mathematics \& Physics Dep. University of Roma Tre, Via Della Vasca Navale, 84, Roma, 00146, Italy
}

\date{Accepted XXX. Received YYY; in original form ZZZ}

\pubyear{2025}

\begin{document}
\label{firstpage}
\pagerange{\pageref{firstpage}--\pageref{lastpage}}
\maketitle

% Abstract of the paper
\begin{abstract}
Before discovering the first exoplanets, the Radial Velocity (RV) method had been used for decades to discover binary stars. Despite significant advancements in this technique, it is limited by the intrinsic mass-inclination degeneracy that can be broken when combining RVs with astrometry, which allows us to determine the orbital inclination, or direct imaging, from which we can estimate the true mass of the target. HD\,72659 is a solar analog known to host a gas giant on a $\sim 10$-yr orbit and a massive outer companion. This work aims to confirm HD\,72659 c, which was recently announced using data from HIRES and HARPS spectrographs in combination with Gaia's astrometric data. We monitored HD\,72659 with HARPS-N in the framework of the GAPS project since 2012. We now combined our 91 spectra with literature data and Gaia DR3 high-precision astrometry to constrain the mass and the orbit of this object ($M_{\rm c} \sim 19$ \mjup, $a \sim 21$ au) that falls in the Brown Dwarf desert. Moreover, we analyzed our high-resolution imaging observation taken with SPHERE, but since the target was not detected, we could only derive upper limits on its mass. We characterize the orbital parameters of HD\,72659 c, confirming the literature mass of this object but finding a period twice as high as previously reported, and we also refine the parameters of planet b with reduced uncertainties compared to previous works. Finally, we analyze and discuss the dynamic configuration of this system, finding that the Kozai-Lidov mechanism may be at work. 
\end{abstract}

% Select between one and six entries from the list of approved keywords.
% Don't make up new ones.

%%%%%%%%%%%%%%%%%%%%%%%%%%%%%%%%%%%%%%%%%%%%%%%%%%

%High-Resolution Echelle Spectrometer (HIRES)
%High Accuracy Radial velocity Planet Searcher (HARPS)
%High Accuracy Radial velocity Planet Searcher for the Northern hemisphere (HARPS-N)
%(Spectro-Polarimetric High-contrast Exoplanet REsearch)
%Global Architecture of Planetary Systems 

%%%%%%%%%%%%%%%%% BODY OF PAPER %%%%%%%%%%%%%%%%%%

\section{Introduction}
For years, the radial velocity (RV) technique was used to detect binary stars \citep[e.g.,][and references therein]{duquennoy1991}. As technology advanced and instrumental precision increased, this allowed us to find exoplanets and brown dwarfs (BDs) as well \citep[e.g.,][]{campbell1988, mayor1995}. Today, we have almost 30 years of high-precision RV data for some stars and are thus beginning to probe the long-orbital-period part of the parameter space (that is, periods larger than a few thousand days) despite several limitations due to well-known selection effects \citep[e.g.,][]{bryan2016}. In particular, long-period planets cause lower RV variations onto their host stars compared to their short-period counterparts of equal mass. In addition, they require systematic monitoring for longer time spans while keeping instruments stable, which is not a trivial task. Lastly, RV suffers from mass-inclination degeneracy and only gives us the minimum mass of the putative companion. For this reason, some of the most massive planets known today might be BDs on nearly face-on orbits. In this context, important help comes from astrometry and direct imaging since these two methods are biased in favor of long-period planets, and allow us to derive the true mass and the three-dimensional orbit of our targets \citep[e.g.,][]{perryman2018}. It is not yet clear at what point a sub-stellar companion must be considered a BD and no longer a planet, with the Deuterium-burning limit being around 13 \mjup\ \citep{saumon2008, spiegel2011}, which is also the reference value used by the IAU Working Group Definition. However, there are objects below this mass limit found in systems likely formed via cloud fragmentation and vice versa, indicating that the formation path is probably a better discriminant than the object's mass. Notorious examples include AB Pic b \cite{chauvin2005} and HD\,168443 \citep{marcy1999,marcy2001,pilyavsky2011}. Nevertheless, in many cases, the mass of the object is used as a discriminant because it is often harder to reconstruct the exact formation and evolution history of a single planetary system, especially if this is old, rather than determining its true mass. A third issue that represents a challenge for the current status of our knowledge is how these massive objects affect the other planets that might be present in the system. In recent years, there has been a lot of effort in searching for new BDs with different techniques, especially in systems with additional planets \citep[e.g.,][]{wilson2016,ryu2017,fontanive2019,kiefer2019,rickman2019,subjak2023,Sozzetti2023}, showing the great interest in the field. This task is made difficult by the estimated low occurrence rates of such objects, which should be around a few percent \citep[e.g.,][]{borgniet2019, nielsen2019, vigan2021}. This phenomenon is known as the "brown dwarf desert", indicating the paucity of this type of companions at low and mid periods. Since the combination of RVs and astrometry yields the true mass of substellar companions and direct imaging, in principle, allows us to derive their spectra, the sum of this information is a great tool to test formation and evolution mechanisms, leading to a better understanding of this phenomenon. Nevertheless, it has been proven that wide BDs do affect the presence and evolution of the other planets in their systems, for example, through the Kozai-Lidov mechanism \citep{fontanive2019}. Many of the mentioned works have been possible thanks to SPHERE at the Very Large Telescope \citep[VLT,][]{sphere}, one of the best direct imaging instruments currently available. Other comparable facilities are mounted at the Keck and Gemini Observatories, leading to the first direct detection of a multiple planetary system around HR 8799 with this technique \citep{marois2008}, and at the Subaru Telescope \citep[e.g.,][]{carson2013}. 

HD\,72659 is a solar-analog star, having the same mass, chemical composition, a mass only 3\% larger, and a similar or larger age (see Section \ref{sec:specparam3}). The star hosts a first planet ($m_{\rm p} \sin{i} \sim 2.8$ \mjup, $P \sim 10$ yr) that was discovered more than 20 years ago by \citet{Butler2003} and its orbit is today very well constrained \citep{Butler2006, Moutou2011}. What's more, \citet{Moutou2011} underline that stellar activity is not correlated with radial velocities and that they find no significant peak in the periodogram of the residuals. However, a long-period signal is visible in our data set, which includes HIRES, HARPS, HRS, and HARPS-N data, covering a time span of about 23 years. Indeed, this trend has also been noted by \citet{bryan2016}, who tried to fit a 2-planets solution. They limited the mass of this new companion to $1.3 \leq M \leq 133$ \mjup\ and the semi-major axis to $7.6 \leq a \leq 35$ au. The values that we retrieved in this work fall in these ranges. \cite{feng2022} announced a second companion for this star, combining GAIA-Hipparcos PMa and RVs (HIRES + HARPS). They reported a mass of 18.81 \mjup, which, as we will show in the next sections, is compatible with our results. However, they reported a period of about 18000 d, much lower than, but formally compatible with, the nominal value that we found from RVs alone. On the contrary, our results from RV + astrometry are not compatible with theirs (see next sections for more details). Motivated by the high mass of this second object and by a previous shallow imaging observation with SPHERE, we decided to observe HD\,72659 again with the same instrument to directly detect its substellar companion, HD\,72659 c. Hosting two massive planets with long orbital periods, this system has the potential to become an interesting benchmark case for dynamic studies. In addition, HD\,72659 has a mass of $\sim 1$ \msun, so the presence of two potentially massive objects was surprising since massive objects require more massive disks, which are typical for higher-mass stars \citep[e.g., ][]{mordasini2012}. 

In Section \ref{sec:obs3}, we describe our RV and astrometric data, plus our new SPHERE observations. In Section \ref{sec:specparam3}, we describe the stellar parameters that we derive from our HARPS-N data. In Section \ref{sec:datatools3}, we describe the tools used in our analysis. In Section \ref{sec:results3}, we describe our results from RV, astrometric, and imaging analysis about the external substellar companion of HD\,72659, followed by a discussion of the dynamics of the system. Finally, in Section \ref{sec:conclusions3}, we summarize our conclusions.

%High-Resolution Echelle Spectrometer (HIRES)
%High Accuracy Radial velocity Planet Searcher (HARPS)
%High Accuracy Radial velocity Planet Searcher for the Northern hemisphere (HARPS-N)
%(Spectro-Polarimetric High-contrast Exoplanet REsearch)
%Global Architecture of Planetary Systems 

\section{Observations and data analysis}
\label{sec:obs3}
\subsection{High-resolution spectroscopy}

\subsubsection{HARPS-N}
We observed HD\,72659 in the framework of the Global Architecture of Planetary Systems (GAPS) programme \citep{covino2013} with HARPS-N (High Accuracy Radial velocity Planet Searcher for the Northern hemisphere) at the Telescopio Nazionale Galileo \citep[TNG,][]{Cosentino2012}. This is a 3.58-m Italian telescope located at the Roque de los Muchachos observatory in the Canary Islands. HARPS-N is a state-of-the-art spectrograph with a resolution of $R = 115000$ that works in the visible part of the electromagnetic spectrum (from 390 to 690 nm). The instrument operates in a vacuum with a temperature stability of $\pm 0.001$ K, ensuring high short- and long-term precision. As part of the Known Planet (KP) sub-program \cite[][Pinamonti et al., in preparation]{benatti2020}, we collected 91 spectra for HD\,72659 from January 3, 2013, to April 11, 2023, observing the target quite intensively between 2013 to 2017, collecting 68 spectra with main goal of searching for inner low-mass planets. Observations were continued at low cadence in the following years with the main goal of characterizing the long-term signal. Radial velocities (RVs) have been extracted using the YABI\footnote{\url{https://ia2.inaf.it/}} online tool \citep{yabi}, which runs the HARPS-N pipeline, using a G2 mask for the cross-correlation function (CCF). The tool also exploits the procedure by \cite{Lovis2011} to derive the Mount Wilson calibrated chromospheric activity index \smw. The median RV uncertainty of our observations is 0.62 m/s. We removed from our data set the observation taken on March 18, 2022, as a clear outlier since the corresponding RV value differs by $\sim 30$ m/s from the observations done in the same period. Moreover, the associated RV uncertainty is 4.37 m/s, which is 7 times larger than the median value, and the signal-to-noise ratio (SNR) in the 6th order ($\lambda = 402$ nm) is 4.80, much lower than the median value of 43.55. The full RV time series is available in Table \ref{tab:hd72659_timeseries} in the Appendix and it is available at CDS via anonymous ftp to \url{cdsarc.u-strasbg.fr} (130.79.128.5)
or via \url{https://cdsarc.cds.unistra.fr/viz-bin/cat/J/MNRAS}.
%The associated \smw\ and BIS (Bisector Inverse Span) values are also rather different from the rest of the data set, confirming that something went wrong during the observation. The reported exposure time is 674 s and the air mass is 1.16 but these two values do not explain what went wrong. However, the SNR in the 6th order is 4.80, while the mean and the median values are 41.75 and 43.55, respectively, confirming that there must have been some technical issue. 

\subsubsection{Archive and literature data}
In addition to our HARPS-N set, we searched the literature for additional data. We used the HIRES (High-Resolution Echelle Spectrometer) data published by \cite{rosenthal2021}, comprising 66 RV data and 47 \smw\ measurements\footnote{The first 19 data points have been taken before the CCD upgrade described in \cite{Butler2017} and, in the data file published, the \smw values are not listed.}. As described by \cite{Butler2017}, HIRES underwent a CCD upgrade procedure in August 2004, so we used different offsets and jitter terms for data taken before and after such an event. 

Furthermore, we downloaded 66 HARPS (High Accuracy Radial velocity Planet Searcher) spectra available in the ESO archive (reduced with the instrument pipeline) and extracted the various activity indices (\smw, BIS, FWHM of the CCF, H$\alpha$, Na I) using the {\small PYTHON} tool ACTIN2 \citep{dasilva2018,dasilva2021}. We removed the first of the two spectra taken on March 22, 2008, as the reported exposure time is only 5 s. We considered separately the data taken before and after the HARPS upgrade described in \cite{LoCurto2015}. 

Finally, we also adopted the data collected with the Hobby-Eberly Telescope (HET) at MacDonald observatory using the High Resolution Spectrograph \citep[HRS,][]{tull1998} instrument and published by \cite{wittenmyer2009}. From these, we removed all the data points with RV uncertainty larger than 10 m/s due to their lower quality compared to the rest of the set, leaving us with 32 of the initial 53 measurements. Table \ref{tab:rv_obs} shows the main characteristics of the RV data sets used in our analysis. 

\begin{table*}%[!htp]
  \caption{Main characteristics of the RV observations of HD\,72659.}\label{tab:rv_obs}
\centering
\begin{tabular}{c c c c c c}
\hline\hline
Data set & N. of obs.  & Start date & End date & Time span [yr] & $\langle \sigma_{RV} \rangle$ [m/s] \\
\hline
HIRES &  66  &  25/01/1998  &  24/10/2019  &  21.74  &  1.44  \\

HARPS &  65  &  23/02/2004  &  05/06/2021  &  17.28  &  1.86  \\

HARPS-N &  90  &  03/01/2013  &  11/04/2023  &  10.27  &  0.74  \\

HRS &  32  &  03/12/2004  &  27/11/2007  &  2.96  &  9.14  \\

Total & 253 & 25/01/1998 & 11/04/2023 & 25.21 & 2.27 \\

\hline
\end{tabular}
\end{table*}

\subsection{Gaia-Hipparcos PMa}

We combined the information obtained from the RV analysis with tangential acceleration information from absolute astrometry, in particular the proper motion anomaly (PMa) from the cross-calibrated Hipparcos-Gaia early data release 3 (EDR3) catalogue of astrometric accelerations by \cite{Brandt2021}. The PMa vector components ($\Delta\mu_\alpha$, $\Delta\mu_\delta$) at the mean epochs of the Hipparcos and Gaia EDR3 catalogues were obtained subtracting from the quasi-instantaneous proper motions of the two catalogues the long-term proper motion vector defined as the catalogues' positional difference divided by the time baseline ($\sim25$ yr). The latter quantity is typically considered as an accurate representation of the tangential velocity of the barycentre of the system in the limit of orbital periods of the companion shorter than the time baseline between the two catalogues. This assumption might be considered questionable given the limited phase coverage of the scaled Hipparcos-Gaia positional difference, spanning only $\sim25$ per cent of the orbital period of HD 72659 c. However, as long as the astrometric acceleration and overall magnitude of the orbital proper motion are small compared to the stellar proper motion (as is our case), the combined analysis of RVs and PMa can still be successfully performed, as demonstrated by a number of recent analyses (e.g., \citealt{Bardalez2021,Currie2023,ruggieri2024b}). 

We note that the reported signal-to-noise ratio of the PMa at the highly sensitive Gaia DR3 mean epoch for HD\,72659 is $\sim 16$, which indicates the presence of a very significant astrometric acceleration. The $\Delta\mu$ values that we used in our RV + astrometry analysis are listed in Table \ref{tab:absastr3}.

\begin{table*}%[ht!]
    \centering
%       \small
        \caption{Proper motion anomaly data for HD\,72659. \label{tab:absastr3}
}
        \begin{tabular}{lcc}
    \hline
    \hline
    \noalign{\smallskip}
    Parameter     &   HD\,72659  &  Reference\\
    \noalign{\smallskip}
    \hline
    \noalign{\smallskip}
    \noalign{\smallskip}
    \noalign{\smallskip}
    HIPPARCOS (epoch 1991.25) $\Delta\mu_\alpha$ (mas yr$^{-1}$) & $+0.059\pm0.690$ & \citet{Brandt2021}\\
        \noalign{\smallskip}
    HIPPARCOS (epoch 1991.25) $\Delta\mu_\delta$ (mas yr$^{-1}$)  & $-1.138\pm0.601$  & \citet{Brandt2021} \\
        \noalign{\smallskip}
    $Gaia$ (epoch 2016.0) $\Delta\mu_\alpha$ (mas yr$^{-1}$) & $+0.311\pm0.043$ & \citet{Brandt2021} \\
        \noalign{\smallskip}
    $Gaia$ (epoch 2016.0) $\Delta\mu_\delta$ (mas yr$^{-1}$)  & $+0.662\pm0.029$&  \citet{Brandt2021} \\
        \noalign{\smallskip}
    \hline
    \end{tabular}

\end{table*}

\subsection{High-contrast direct imaging}
\label{sec:imagingdata}
%\begin{figure}
%\vspace{-2cm}
%   \centering
%   \includegraphics[width = \linewidth]{Figures/old_sphere.png}
%      \caption{Archive SPHERE observation of HD\,72659. The blue circle represents our tentative detection with SNR = 6.
%              }
%         \label{fig:old_sphere}
%   \end{figure}
Our target was observed with SPHERE at VLT \citep{sphere} on December 31, 2015, during the GTO survey SHINE \citep[ESO ID: 096.C-0241(C),][]{desidera2021}. The observation was part of a filler program aiming at stars hosting RV-discovered exoplanets (Hagelberg et al., in prep.). The data were reduced through the SPHERE Data Center \citep{spheredatacenter}, exploiting the instrument pipeline \citep{spheredrh} and the SpeCal procedures \citep{specal}. Considering the relatively shallow detection limits due to the short integration time and small field rotation, and the fact that the mass of the companion was likely high due to the RV trend and the high-SNR astrometric acceleration, we applied for a new and deeper observation. 
%We retrieved a weak (SNR = 6) potential detection at the same expected position for a companion responsible for the PMa, as shown in Figure \ref{fig:old_sphere}. Motivated by this, we applied for a new and deeper observation. \\

HD\,72659 was observed again with SPHERE following our accepted proposal (ESO ID: 112.25JY, PI: A. Ruggieri) on the night of January 21, 2024, exploiting the IRDIFS\_EXT observing mode. In this mode we observed both with IFS \citep{ifs} covering $Y$, $J$, and $H$ spectral bands between 0.95 and 1.65 microns and with a field of view (FOV) of $1.7^{\prime \prime} \times 1.7^{\prime \prime}$, and with IRDIS \citep{irdis} operating in the K spectral band using the K12 filter pairs \citep[wavelength K1 = 2.110 microns; wavelength K2 = 2.251 microns;][]{vigan2010} on a circular FOV of $\sim 5^{\prime \prime}$. The observations were performed exploiting the SPHERE extreme adaptive optics system SAXO \citep{fusco2006}. Before and after the science coronagraphic data, we also acquired frames with the stellar PSF offset with respect to the coronagraph to perform photometric calibration. To this aim, we used a neutral density filter to avoid saturation of the detector. Moreover, we also acquired frames with satellite spots at symmetrical positions with respect to the central star to precisely define the position of the star behind the coronagraph \citep{marois2012}. The science data were acquired in pupil-stabilized mode, allowing the rotation of the FOV during the acquisition to be able to implement the angular differential method \citep[ADI, ][]{marois2006}. In total, we acquired 48 frames with a detector integration time (DIT) of 64 s for a total exposure time of 3072 s. The total rotation of the FOV during the observation was 28.49 degrees. The data were reduced using the SPHERE data reduction and handling pipeline \citep[DRH,][]{pavlov2008}. Finally, we performed speckle subtraction to reduce the noise from the star and highlight the signal from the star at the same time. To this aim, we applied at the same time both ADI and spectral differential imaging \citep[SDI,][]{racine1999}, implementing the principal components analysis \citep[PCA,][]{soummer2012}. Table \ref{tab:sphere_obs} contains the technical information for both the old and new SPHERE observations.

\begin{table*}%[!htp]
  \caption{Main characteristics of the SPHERE observations of HD\,72659.}\label{tab:sphere_obs}
\centering
\begin{tabular}{ccccccccc}
\hline\hline
Date  &  Obs. mode & Coronograph & DIMM seeing & $\tau_0$ (ms) & wind speed (m/s) & Field rotation & DIT & Total exposure\\
\hline
2015-12-31  & IRDIFS & N\_ALC\_YJH\_S & 0.72" & 3.0 & 3.58 & 9.6$^{\circ}$ & 64 s & 1024 s \\
2024-01-21  & IRDIFS\_EXT & N\_ALC\_YJH\_S & 0.72" & 7.8 & 7.60 & 28.49$^{\circ}$ & 64 s & 3072 s \\

\hline
\end{tabular}
\end{table*}

\section{Stellar parameters}
\label{sec:specparam3}

\subsection{Spectroscopic parameters}
We derived spectroscopic stellar parameters for HD\,72659 using HARPS-N spectra, employing the same procedure described in \cite{ruggieri2024b}. Through equivalent width (EW) measurements and the force-fitting abundance method, we adopted the \texttt{q}2 {\small PYTHON} code \citep{ramirez2014}, allowing the use in silence mode of the 2019 version of the MOOG radiative transfer code \citep{1973Sneden}. Using the \cite{2003castelli} grid of model atmospheres and the line list by \cite{2020dorazi}, effective temperature and surface gravity were derived via excitation/ionization equilibrium of FeI and FeII lines. Microturbulence values were obtained by minimizing the trend between the reduced EW of Fe I lines and the corresponding iron abundance. The outcome of this standard procedure gives us the final iron abundance. Errors in stellar parameters were evaluated using q2, considering dependencies among parameters. Internal errors on metallicity result from the quadrature sum of errors from EW measurements and uncertainties related to stellar parameters. Then, we measured the projected rotational velocity $v\sin i$ using the spectral synthesis technique and the iSpec package (\citealt{BlancoCuaresmaetal2014}), where stellar parameters were fixed to those previously derived (see \citealt{ruggieri2024b} for details on the procedure). The star results to be a slow rotator (1.6$\pm$0.9 \kms). The corresponding inferred rotation period is highly uncertain, $P_{rot} = 43.0 \pm 24.3$ d.  Another estimate of $P_{rot}$ can be obtained from \logrhk\ value\footnote{For the estimate of the expected rotation period, we used the  \logrhk\ derived from Sindex using the B-V color of the star (0.612 from Hipparcos catalog), which results \logrhk\=-5.00 to be consistent with the formalism by \cite{mamajek2008}. In Sect. \ref{sec:age} we will instead adopt the logrhk value derived following \cite{oliveira2018}, which is more suitable to our target star.} finding $20.7 \pm 1.3$ d \citep[using the expression by][]{noyes1984} and $21.9 \pm 1.9$ d \citep[using the expressions by][]{mamajek2008}. From our adopted age value in Sect. \ref{sec:age}, the expected rotation period is 24 and 32 d for \cite{mamajek2008} and \cite{ye2024}, respectively. These values and especially their uncertainties should be taken with caution, considering the ambiguities in the rotational evolution of old solar-type stars \citep{vansaders2016,hall2021}. In any case, it is plausible that the rotation period of HD 72659 is in the range 20 - 30 d. Table \ref{tab:starparam3} shows the results for all stellar parameters.

%In the end, we estimated the rotation period from \logrhk\ values, finding $22.37 \pm 1.31$ d \citep[using the expression by][]{noyes1984} and $24.40 \pm 1.85$ d \citep[using the expressions by][]{mamajek2008}. 

%However, these values are likely too low for a star much older than the Sun (see the isochrone fitting part in the next Section). What's more, \cite{mamajek2008} reported that their rotation-stellar activity relationship breaks down for less active stars, as is our case, in agreement with previous results by \cite{wright2004}. Therefore, we estimated the rotation period using the \vsini\ value calculated above. In particular, we find $P_{rot} = 43.04 \pm 24.29$ d and notice that this is not very informative since the relative error is about 50\%. This is caused by the large uncertainty on $v \sin i$ as we are close to the resolution limit. Nevertheless, this higher value of rotation period, despite the large uncertainty, is more reliable than the one derived using the \logrhk\ indicator. 

\subsection{Stellar age}
\label{sec:age}

\begin{figure}
%\vspace{-2cm}
   \centering
   \includegraphics[width = \linewidth]{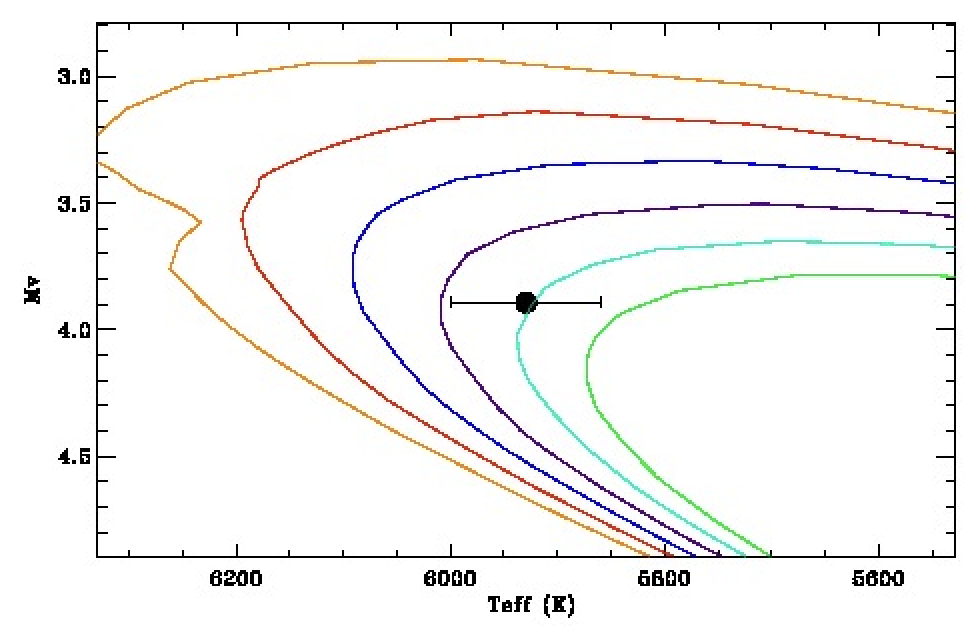}
      \caption{Isochrone fit for HD\,72659. PARSEC isochrones of metallicity [Fe/H]=-0.02 and ages 4.0,5.0,6.0,7.0,8.0, and 9.0 Gyr are shown.
              }
         \label{fig:isoc_kp59}
   \end{figure}

We used different methods to derive the age of HD\,72659. First, because HD\,72659 evolved outside the main sequence, we can determine its mass and age well through isochrone fitting. Using PARSEC isochrones \citep{parsec} and the metallicity derived from HARPS-N spectra ([Fe/H] $= -0.02 \pm 0.06$), we inferred an age of $8.0 \pm 1.0$ Gyr and, thus, a mass of $1.033 \pm 0.025$ \msun. Figure \ref{fig:isoc_kp59} shows the result of the isochrone fitting. Secondly, since the target spectrum shows the presence of the lithium line at $\sim$6707.8\,$\Angstrom$, we estimated from the Li EW an abundance compatible with the age of the M67 open cluster ($\sim$4.5 Gyr, \citealt{Pasquinietal2008}), where correction by \cite{Lindetal2009} for NLTE effects was considered. Next, following the procedure by \cite{Biazzoetal2022}, we derived the abundance of the $s$-process yttrium element and the $\alpha$ magnesium and aluminum elements, as ratios [Y/Mg] and [Y/Al] are considered to be helpful chemical clocks, in particular for solar analogs (\citealt{casalietal2020}). Thanks to these elemental ratios, we derived a mean age of $6.2 \pm 1.0$ Gyr (see \citealt{ruggieri2024b} for details). Successively, we estimated the star's age from the value of the Ca II H\&K activity index using the equations by \cite{oliveira2018}, which are specifically valid for solar twins. In particular, using the value of \teff\ and \smw\ in Table~\ref{tab:starparam3}, we obtained $\rm log\,R^{\prime}_\mathrm{HK} (T_{\mathrm{eff}})= -5.04 \pm 0.025$, from which we calculated an age of $5.4 \pm 1.2$ Gyr. We notice that this value of $\log\,R^{\prime}_\mathrm{HK}$ is slightly different from the value in Table~\ref{tab:starparam3} because the correction for the photospheric flux contribution is evaluated as a function of \teff\ and not $B-V$ as usual \citep[cf.][]{oliveira2018}. The X-ray emission does not provide significant constraints, as the upper limit the X-ray luminosity by \cite{kashyap2008} corresponds
to an age older than 1.2 Gyr using \cite{mamajek2008} calibration.
Finally, we used the technique proposed in \cite{almeida2018} to calculate the age probability distribution from the U, V, and W components of spatial motion. The expected age according to equation 16b of the authors is 7.11 Gyr, the peak of the distribution is at 4.96 Gyr, and the kinematical age (according to the definition of the same authors) is 5.50 Gyr. For all these estimates, the uncertainty is 3 Gyr. Among all these methods, we consider elemental ratios and isochrones the most reliable since HD\,72659 is very similar to the Sun in mass and chemical composition. In addition, age estimates from kinematics suffer from large uncertainties when applied to single stars and those from chromospheric activity are affected by magnetic cycles and rotation, which is dubious in our case. Considering the age obtained from the isochrone fitting and the elemental ratios, we find that the two are consistent with each other at $1.3\sigma$. In the following, we adopt the age obtained from the isochrones, although we note that, due to a partial discrepancy with the nominal values obtained with the other methods, the true age could likely be in the lower part of the error bar. 
%\begin{figure}
%\vspace{-2cm}
%   \centering
%   \includegraphics[width = \linewidth]{Figures/age_pdf.pdf}
%      \caption{Age probability distribution calculated using the method in \protect\cite{almeida2018}.}
%         \label{fig:age_pdf}
%   \end{figure}
   
\begin{table}
\centering\caption{Stellar parameters of HD\,72659}
\begin{tabular}{l c c }
\hline
\hline
Parameter & Value  & Ref. \\
\hline
Parallax~[mas]  & $19.258 \pm 0.027$ & 1\\
Spectral type & G2V & 2 \\
$V$ [mag] & 7.46 & 2 \\
$B-V$ [mag] & 0.612 & 2 \\
$\mu_{\alpha^{\star}}$~[mas\,yr$^{-1}$] & $-112.313	\pm 0.026$ & 1 \\
$\mu_{\delta}$~[mas\,yr$^{-1}$] & $-96.386	\pm 0.018$ & 1 \\
U\ [\kms]      & $7.17 \pm 0.11$ & 3 \\
V\ [\kms]      & $-2.04 \pm 0.12$ & 3 \\
W\ [\kms]      &  $-39.99\pm0.09$ & 3 \\
ecc (galorb)   & 0.07 & 4 \\
zmax [kpc]     & 0.59 & 4  \\  
RV [\kms]      & $-18.203_{-0.029}^{+0.024}$ & 5 \\
\teff\ (spec.) [K]  & $5929 \pm 61$  &  5 \\
\logg\ [cgs]  &  $4.23 \pm 0.14$ & 5 \\
\feh\ [dex]   & $-0.02 \pm 0.06$ & 5 \\
$\xi$ [\kms]  &  $1.15 \pm 0.08$ & 5 \\
\mstar\ [$\mathrm{M_\odot}$] & $1.033 \pm 0.025$ & 5 \\ %%
\rstar\ [$\mathrm{R_\odot}$] & $1.36 \pm 0.06$ & 5 \\
Age [Gyr]  &  $8.0 \pm 1.0$  & 5 \\
\vsini\ [\kms]  & $1.6 \pm 0.9$ & 5 \\
%\prot\ from \logrhk\ [d]  & $20.7 \pm 1.3$, $21.9 \pm 1.9$ & 5 \\
%\prot\ from \vsini\ [d]  & $43.04 \pm 24.29$ & 5 \\
S$_{\text{MW}}$ (HARPS-N)  & $0.157 \pm 0.003$ & 5 \\
\logrhk\ (HARPS-N) & $-5.00 \pm 0.02$ & 5 \\
$EW_{\rm Li}$ [\AA]  & $36.5 \pm 1.0$ & 5 \\
A(Li)$_{\rm NLTE}$ & $2.25 \pm 0.05$ & 5 \\
$\log L_{X}$  &  $< 28.78$ & 6 \\ 
%$\log L_{X}/L_{bol}$  &  $< 28.78$ & 6 \\ 
\hline
\end{tabular}
\label{tab:starparam3}
\vskip0.2cm
References: 1: \textit{Gaia} DR3; 2: \citet{Moutou2011}; 3: \citet{smart2021}; 4: \citet{mackereth2018}; 5: this work; 6: \citet{kashyap2008}
\end{table}

\section{Data analysis}
\label{sec:datatools3}
\subsection{Periodograms}
We used the \texttt{astropy} {\small PYTHON} package to extract the GLS \citep[Generalized Lomb-Scargle periodogram,][]{gls} of our RV time series (including both old and new data). In the following, we first extracted the GLS to each data set individually to account for the different offsets and then applied it to the whole data set after subtracting each offset from the corresponding individual set. We calculated False Alarm Probabilities (FAPs) using the implemented bootstrap method. In practice, the periodogram is calculated repeatedly on many resamplings, each time keeping the temporal coordinates unchanged, while the RV values are drawn randomly from the observed values. See the \texttt{astropy} documentation for more details.

\subsection{RV orbital fit}
For the RV analysis in this work, we used the 9.1 version of \texttt{PyORBIT}\footnote{\url{https://pyorbit.readthedocs.io/}} \citep{Malavolta2016, Malavolta2018} to perform a Monte Carlo Markov Chain (MCMC) analysis of our RV data. To do this, the program exploits the global optimization algorithm \texttt{PyDE}\footnote{\url{https://github.com/hpparvi/PyDE}} \citep{storn1997} and the affine invariant MCMC sampler \texttt{emcee}\footnote{\url{https://emcee.readthedocs.io/en/stable/}} \citep{foreman13}. We applied different offset and jitter terms to data taken with different instruments. For each planet, the code fits the data using a Keplerian curve with the following parameters: the RV semi-amplitude $K$, the orbital period $P$, the mean longitude $L = \omega + M_0$ \citep[setting $\Omega=0$ as shown in Sect. 4.2 in][]{ford2006}, $\sqrt{e}\sin\omega$, and $\sqrt{e}\cos\omega$, where $\omega$ is the argument of periastron of the planet and $e$ the orbital eccentricity \citep[following the parametrization by][]{Eastman2013}. The semi-major axis and the minimum mass ($m\sin i$) are then derived using the stellar mass. We used uniform priors for all parameters considered. In total, we have 22 parameters to be fitted: 6 offset terms, 6 jitter terms, and 5 parameters for each planet, giving a total of 88 walkers. We ran the algorithm for 200k steps and verified that chains have converged with the integrated autocorrelation analysis \citep{goodman2010}, the Gelman-Rubin diagnostic using a reference value of 1.01 \citep{gelman1992}, and visual inspection. As explained in Section \ref{sec:rv_fit_3}, we fit the data with different models and then compared them using the Bayesian Information Criterion \citep[BIC,][]{schwarz1978} value and the results by \cite{kass1995}.

\subsection{Combined fits: RVs and astrometry}\label{sec:pma_rv_fit3}
We combined our RV results with astrometric data with the DE-MCMC algorithm \citep{TerBraak2006,Eastman2013} described in \cite{Sozzetti2023}, following the same procedure as in \cite{ruggieri2024b}. This allows us to derive the values of the orbital inclination $i$, the longitude of the ascending node $\Omega$, and the mass ratio $q$. We adopted uninformative priors for the model parameters derived by the RV-only analysis and uniform priors on $\cos(i)$, $\Omega$, and $q$.

\section{The substellar companion of HD\,72659}
\label{sec:results3}
%\subsection{Previous works}

\subsection{Periodogram analysis} 
\begin{figure*}
   \centering
\includegraphics[width = 0.5\linewidth]{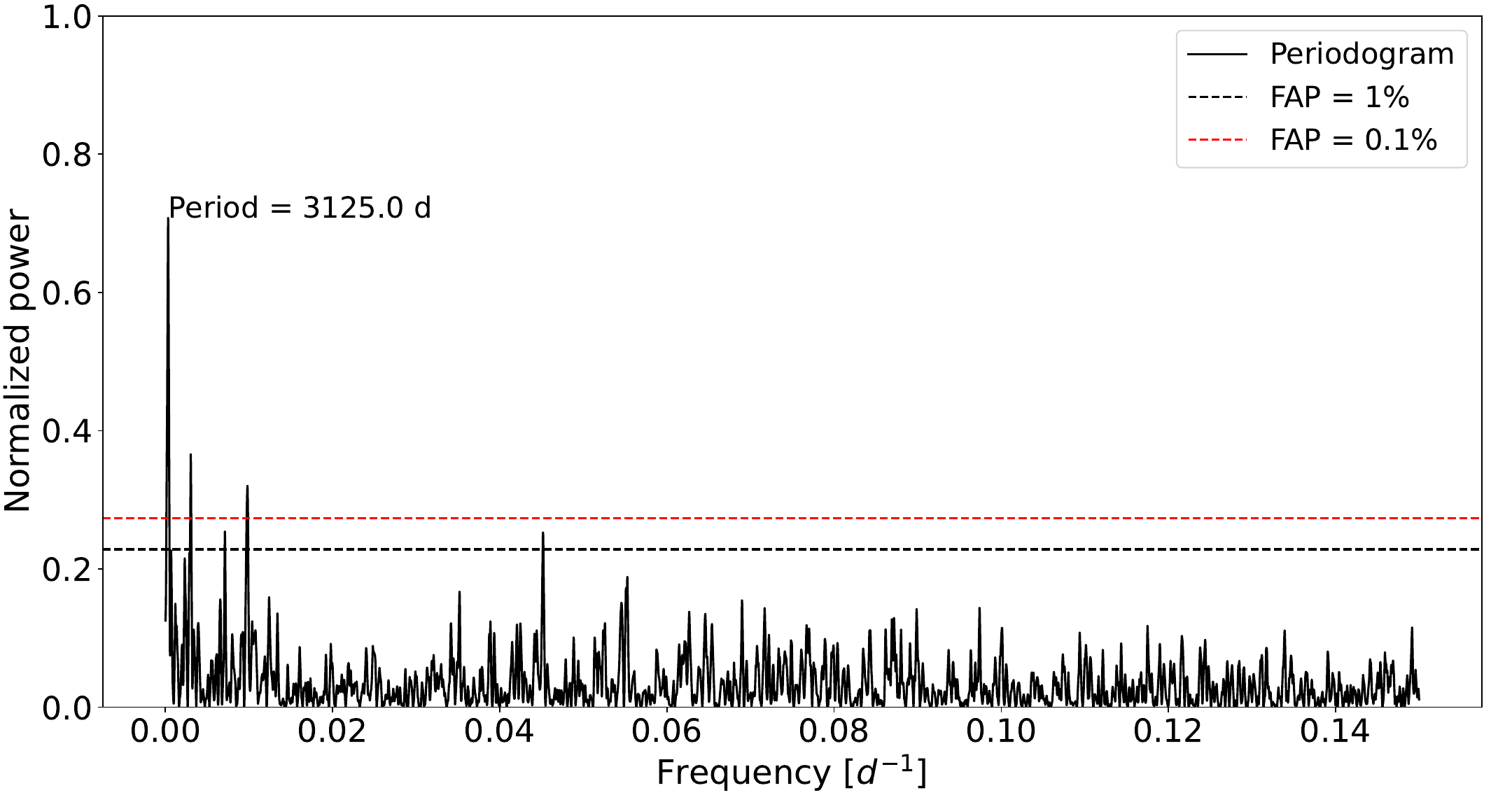}\includegraphics[width = 0.5\linewidth]{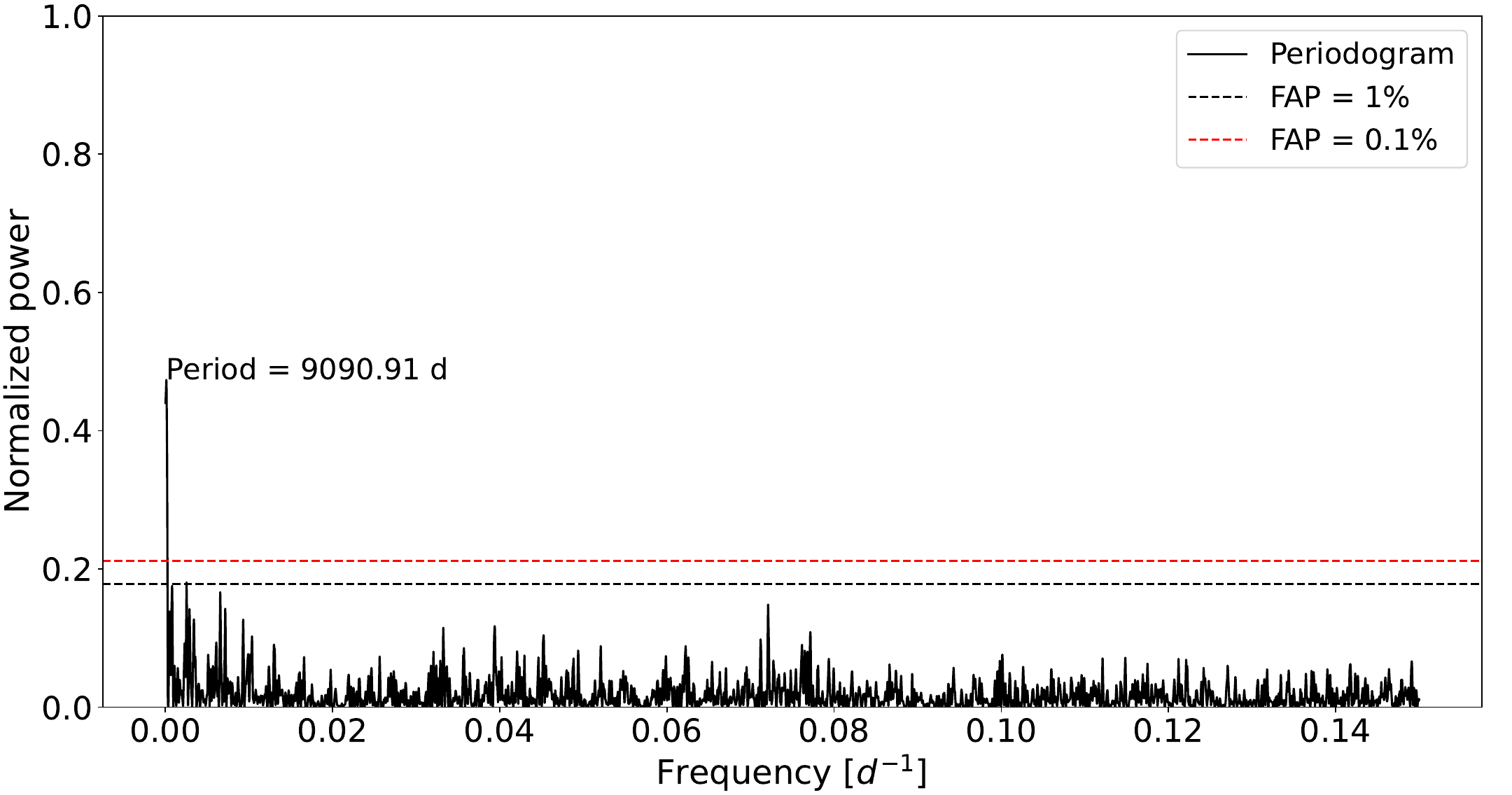}
      \caption{GLS of RV data of HD\,72659, before (left) and after (right) subtracting the signal from planet b. 
              }
         \label{fig:gls_rv}
   \end{figure*}

\begin{figure*}
   \centering
\includegraphics[width = \linewidth]{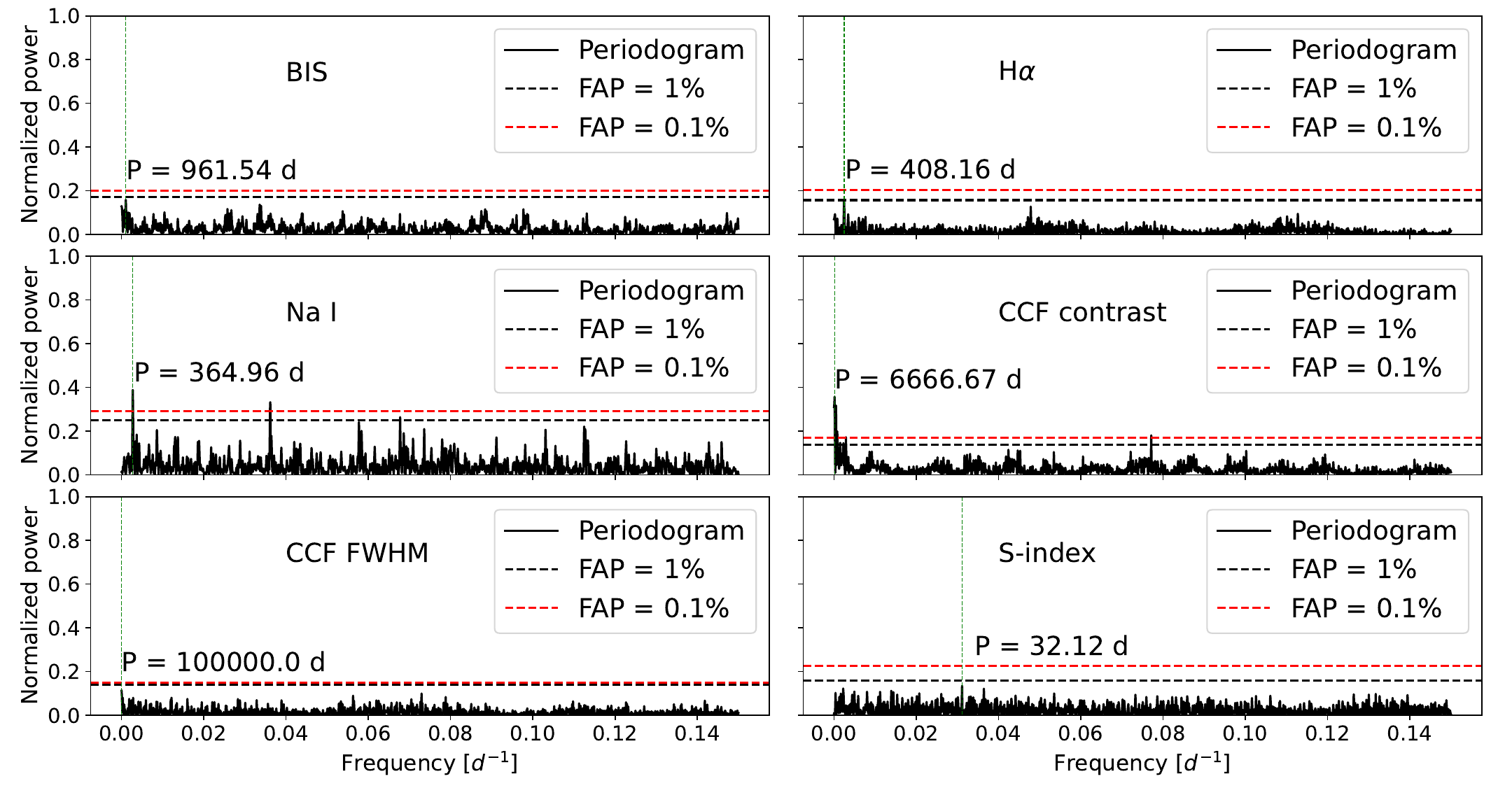}
      \caption{GLS of activity data of HD\,72659. 
              }
         \label{fig:gls_activity}
   \end{figure*}

As shown in Figure \ref{fig:gls_rv} (left panel), in the GLS of the RVs of HD\,72659, we find the peak corresponding to planet b to be extremely clear, even though at slightly lower periods, probably because the GLS pipeline fits a sinusoidal curve with no eccentricity. In the residuals, after fitting a simple sinusoidal with the automatic pipeline at the nominal peak found by the previous GLS, we can see a longer-term signal (Figure \ref{fig:gls_rv}, right panel). The nominal peak is at a much shorter period than later found with the full RV analysis, again likely due to the eccentricity matter and because the time span of the observations is much shorter than the real period of planet c. As far as the activity is concerned, we extracted the GLS of the \smw, BIS, CCF contrast, CCF FWHM, H$\alpha$, and Na I activity indicators \citep[for HARPS data, the \smw, H$\alpha$, and Na I indices have been derived from the spectra using the tool ACTIN2 by][]{dasilva2021}. All the periodograms for these activity indices are shown in Figure \ref{fig:gls_activity}. As we can see, the only formally meaningful peaks are found in the sodium and CCF contrast time series. We therefore checked whether these data showed any correlations with RVs. Considering Na I, the Spearman's rank coefficient is $\abs{r}{} < 0.1$ both for HARPS and HARPS-N data, indicating no correlation. For the CCF contrast, we obtained $r = 0.285$ for HARPS and $r = -0.447$ for HARPS-N, indicating a possibly weak anti-correlation in the second case. However, as described in \cite{benatti2017}, during the first part of the observations the instrument was affected by a small defocus that was fixed in March 2014. Thus, removing the data taken before that date, we obtain for the HARPS-N data set $r = 0.137$, once again indicating no correlation between RVs and the CCF contrast. We, therefore, confirm that there does not seem to be neither rotational modulations nor any significant activity cycle that might affect the RV time series. 

\subsection{RV orbital fit}
\label{sec:rv_fit_3}
%\begin{figure*}
%   \centering
%   \includegraphics[width = \linewidth]{Figures/HD72659_kep_lin_quad.pdf}
%      \caption{Keplerian, linear, and quadratic models for the RVs of HD\,72659 subtracted planet b.
%              }
%         \label{fig:kp59models}
%   \end{figure*}
For our analysis, we used six different offsets and jitters: HIRES pre-upgrade, HIRES post-upgrade, HARPS-N, HARPS pre-upgrade, HARPS post-upgrade, and HRS. We fitted our data with a one-planet model (Model 1, M1), a 2-planets model (Model 2, M2), and one planet plus linear (Model 3, M3) or quadratic trend (Model 4, M4). We then compared the results from these based on the BIC criterion by \cite{kass1995} and derived lower limits for the RV semi-amplitude $K$, orbital period $P$, and minimum mass $m\sin i$ of the outer planet using our polynomial coefficients and the equations by \cite{kipping2011}. \\

The results obtained with all these models for HD\,72659 are shown in Table \ref{tab:tab_compare_models}, where we also reported the BIC value for the model with only planet b for comparison. As we can see from the Table, when using a Keplerian term for the outer planet (M2) in this system, we found a highly unconstrained period as our data cover $\sim 1/3$ of the nominal value. Nevertheless, we note that our result agrees with the one by \cite{feng2022} within $1.1\sigma$. When using a linear polynomial instead (M3), we found smaller nominal values for $K$, $P$, and $m\sin i$, but since these are just lower limits, they are in overall agreement with the previous model. In contrast, when using a second-order polynomial, we obtained a much longer orbital period that is incompatible at a $> 2.5\sigma$ level, even though the $m\sin i$ is compatible with our previous estimates. As we can see from the BIC values shown in Table \ref{tab:tab_compare_models}, the one-planet model is by far the worst at fitting the data. Adding a second Keplerian significantly improves the fit and is favored over a first-order polynomial, indicating that a curvature is present in our data set. However, the quadratic trend is the most favored since we only see a relatively small fraction of the orbit of this object. \\

\begin{figure}
%\vspace{-2cm}
   \centering
   \includegraphics[width = \linewidth]{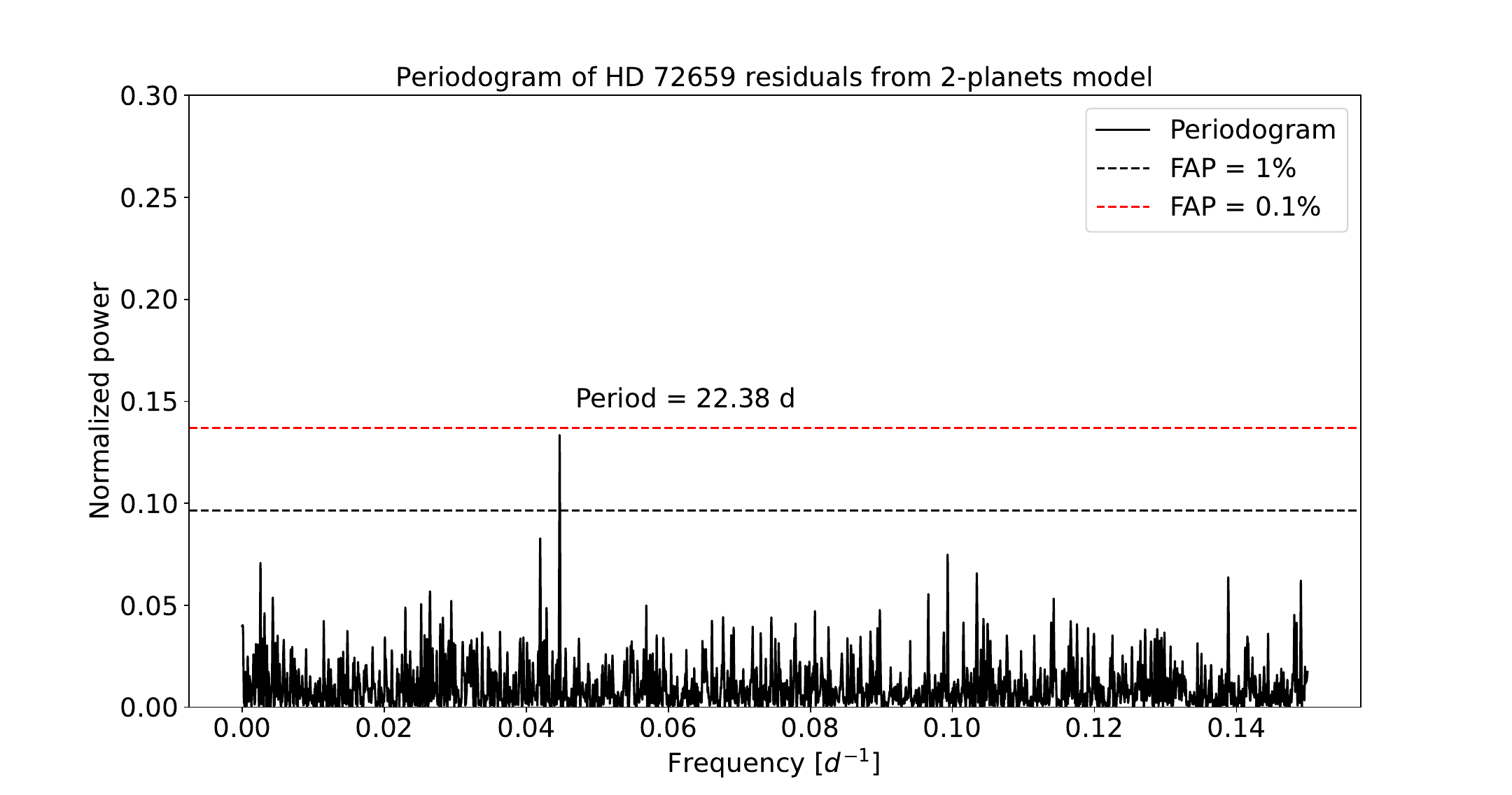}
      \caption{GLS of the RV residuals for HD\,72659 after removing the signals corresponding to planets b and c.
              }
         \label{fig:hd72659_gls}
   \end{figure}

\begin{figure}
%\vspace{-2cm}
   \centering
   \includegraphics[width = \linewidth]{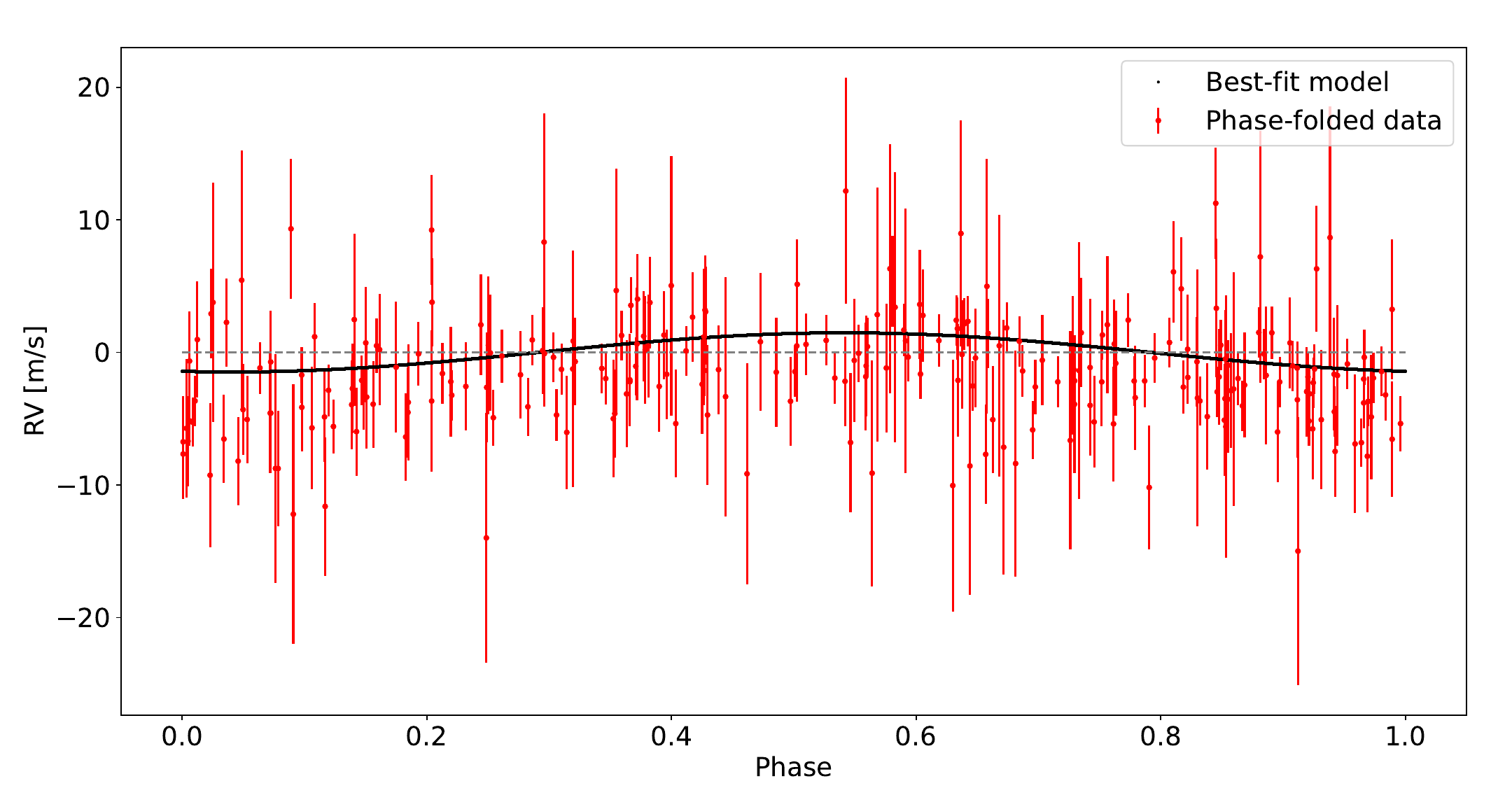}
      \caption{Phase-folded data and Keplerian model for the 22-d signal present in the residuals.
              }
         \label{fig:phase_d}
\end{figure}

\begin{figure}
   \centering
   \includegraphics[width = \linewidth]{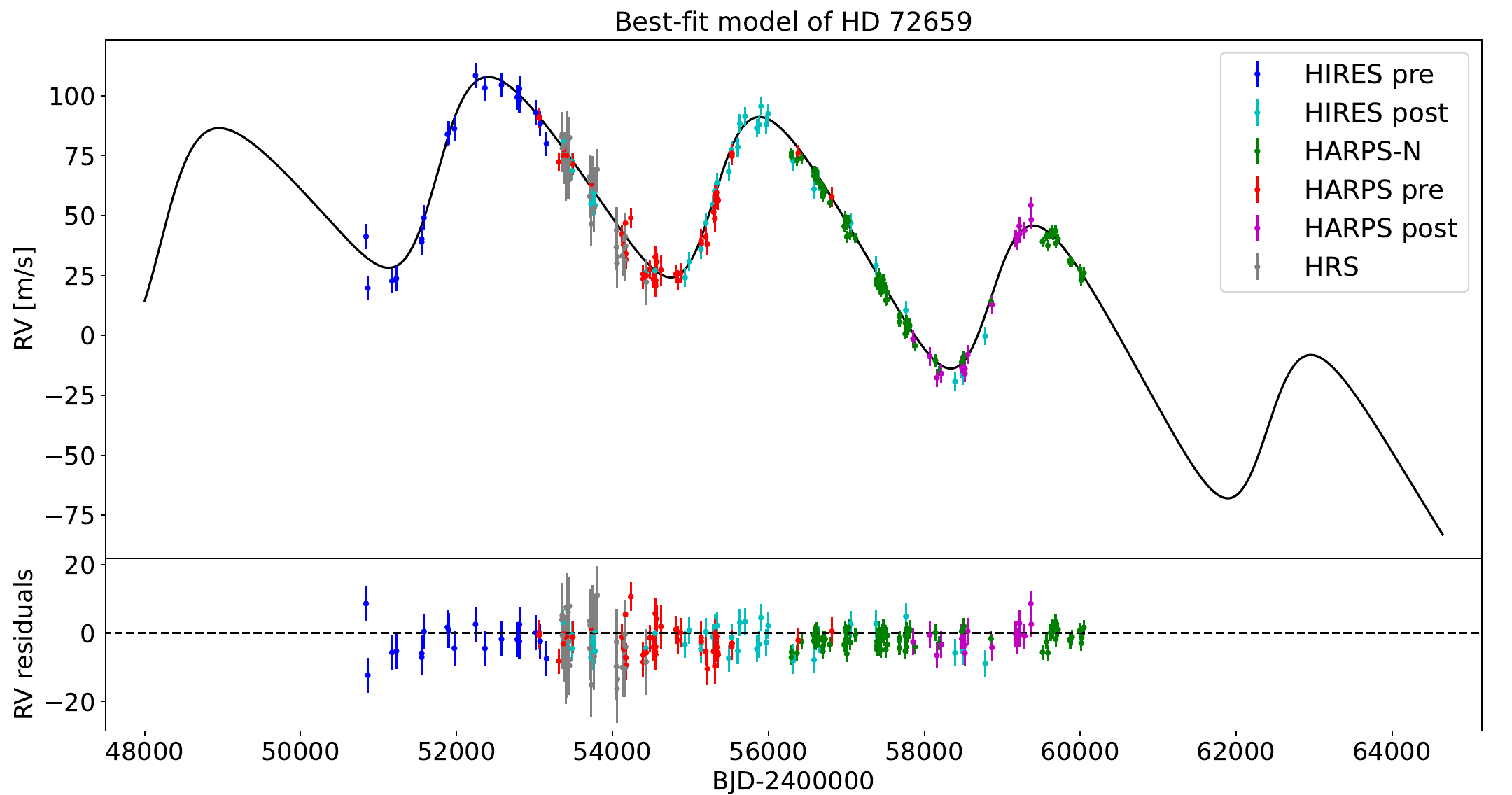}
      \caption{Best-fit model for HD\,72659 obtained using two Keplerian terms.
              }
         \label{fig:kp59bestfit}
   \end{figure}

We conclude that RVs alone are not enough to accurately determine the orbital and physical properties of this object and thus resort to a combined RV + astrometry analysis. Before doing so, we searched for additional signals in the residuals of the quadratic and 2-planet models, finding in both cases a significant (FAP $\sim 0.1\%$) peak at 22 d, as shown in Figure \ref{fig:hd72659_gls}. Thus, we repeated the analysis, fitting the data with three Keplerian terms, setting a flat prior for the period of the third one in the range $[2, 100]$ d and assuming a circular orbit. The results for planets b and c are consistent with the previous models, while the third Keplerian corresponds to a $6.7 \pm 1.3$ \mearth\ planet with $P_d = 22.3754 \pm 0.0083$ d. This model is statistically favored over the 2-planets one with $\Delta\text{BIC} = 53$. However, we refute this putative third planet based on the following considerations. First of all, we extracted the GLS of the S-index time series and found the highest peaks for HARPS-N and HARPS data at 29 and 22 d, respectively, although both have FAPs $> 1\%$. No such signal is present in the HIRES data as this is the most sparse set. In any case, despite the low significance of these peaks, there is an indication that the 22 d signal seen in the RVs might be related to stellar activity, considering that these periodicities are in the range of the expected rotation period for the star estimated in Sect. \ref{sec:specparam3}.
% frase tolta dopo la discussione con Nuccio
%In particular, we note that the 22-d value found here is exactly one half of the rotation period estimated from the \vsini. A signal with a period corresponding to the first harmonic of the rotation period is what one can expect when the RV variation is induced by active regions perturbing the photospheric flux, like spots and/or faculae \citep[see Sect. 3.2.1 in ][]{lanza2010}. 
%Secondly, 
Furthermore, the jitter terms associated with each RV data set do not diminish when adding a third Keplerian to the fit and the same holds for the uncertainties on the orbital parameters of planets b and c. If this was a real signal, we would expect them to decrease, while in our case they are identical to the 2-planets case. What's more, the jitters range from 1.8 to 5 m/s while $K_d = 1.48 \pm 0.29$ m/s, meaning that the signal has a lower amplitude than the jitters. On top of that, as shown in Figure \ref{fig:phase_d}, the phase-folded model for this presumed object does not follow the data as well as one would expect. Considering all this, we do not further consider the presence of a third planet in the rest of the paper and leave a potentially more detailed analysis for future work. The results obtained with the two Keplerian terms are listed in Table \ref{tab:tabparam3} and the best-fit model is shown in Figure \ref{fig:kp59bestfit}. 

\begin{table*}
%%\small
\centering
\caption{\label{tab:tab_compare_models} Comparison between the various models used to fit RV data}
\begin{tabular}{ c c c c c c c c }
\hline
\hline
\\
 Model & N. of planets & $\dot{\gamma}$ [ms$^{-1}$d$^{-1}$] & $\ddot{\gamma}$ [ms$^{-1}$d$^{-2}$] & $K_c$ [m/s] & $P_c$ [d] & $m_c \sin i_c$ [\mjup] & BIC \\
\\
\hline
\\
M1 &  1  &  -  &  -  &  -  &  -  &  -  & 1917 \\
\\
\hline
\\
M2 &  2  &  -  &  -  &  $74_{-23}^{+28}$ &  $27960_{-8683}^{+9834}$ &  $11.1_{-4.1}^{+5.5}$  & 1574 \\
\\
\hline
\\
M3 &  1  &  $(-1.267 \pm 0.031) \cdot 10^{-2}$  &  -  & $\geq 45.52$  &  $\geq 14372$  &  $\geq 5.34$  & 1593 \\
\\
\hline
\\
M4 &  1  &  $\left( -2.70_{-0.67}^{+0.64} \right) \cdot 10^{-3}$  &  $\ddot{\gamma} = (-1.016 \pm 0.068) \cdot 10^{-6}$  &  $\geq 71.26$  &  $\geq 52622$  &  $\geq 12.88$  & 1549 \\
 \\
\hline
\end{tabular} \\
\vspace{0.2cm}
Linear coefficient, quadratic coefficient, RV semi-amplitude, orbital period, and minimum mass derived for HD\,72659 c using different models. The second column lists the number of planets included in the model, while the last one shows the corresponding BIC values. 
\end{table*}

%\hskip-2cm
\begin{table}
%%\small
\centering
\caption{\label{tab:tabparam3} Planetary parameters obtained from the RV analysis using two Keplerian terms}
\renewcommand{\arraystretch}{0.6}
\begin{tabular}{ c c c }
\hline
\hline
\\
 Parameter & HD\,72659 b & HD\,72659 c \\
\\
\hline
\\
$m_{\rm p} \sin{i}$ [\mjup] & $2.851 \pm 0.057$ & $11.1_{-4.1}^{+5.5}$ \\
\\
\hline
\\
P [d] & $3549 \pm 10$ & $27960_{-8683}^{+9834}$ \\
\\
\hline
\\
a [a.u.] & $4.607 \pm 0.039$ & $18.3_{-4.0}^{+4.1}$ \\
\\
\hline
\\
e & $0.2393 \pm 0.0098$ & $0.104_{-0.074}^{+0.14}$ \\
 \\
\hline
\\
$\omega$ [degrees] & $-95.9 \pm 2.6$ & $-112_{-35}^{+212}$ \\
\\
\hline
\\
K [m/s] & $38.23 \pm 0.46$ & $74_{-23}^{+28}$ \\
\\
\hline
\\
$T_0$ [BJD - 2450000] & $3594 \pm 22$ & $10299_{-1582}^{+1674}$ \\
\\
\hline
\\
jitter HARPS-N [m/s] & \multicolumn{2}{c }{$2.07_{-0.18}^{+0.20}$} \\
\\
\hline
\\
jitter HIRES [m/s] & \multicolumn{2}{c }{$4.86_{-0.93}^{+1.20}$} (pre-upgrade) \\
\\
 & \multicolumn{2}{c }{$3.66_{-0.44}^{+0.53}$} (post-upgrade) \\
\\
\hline 
\\
jitter HARPS [m/s] & \multicolumn{2}{c }{$3.53_{-0.49}^{+0.56}$} (pre-upgrade) \\
\\
& \multicolumn{2}{c }{$3.58_{-0.69}^{+0.90}$} (post-upgrade) \\
\\
\hline
\\
jitter HRS [m/s] & \multicolumn{2}{c }{$2.0_{-1.4}^{+2.1}$} \\
\\
\hline
\end{tabular} \\
\vspace{0.2cm}
Physical and orbital parameters for the planets of HD\,72659 derived from RV fitting. The reported values are the medians of the resulting distributions and the error bars are the 15th and 84th percentile, respectively. The parameter $T_0$ is the time of the periastron passage calculated with reference time BJD $= 2453000$.
\end{table}

\subsection{Orbital parameters and true mass from combined fits}

\begin{table}%[ht!]
    \centering
%       \small
        \caption{Orbital parameters and true mass for HD\,72659 c from the combined RV+astrometry analysis. \label{tab:fit_HD72659}
}
        \begin{tabular}{lcc}
    \hline
    \hline
    \noalign{\smallskip}
    Parameter     &  Prior &  Value\\
    \noalign{\smallskip}
    \hline
    \noalign{\smallskip}
    \noalign{\smallskip}
    $P_c$ [yr] & $\mathcal{U}(30.0,150.0)$  & $97.1^{+3.4}_{-2.5}$  \\
    \noalign{\smallskip}
    $T_{0,c}$ [yr] & $\mathcal{U}(0.0,3000.0)$  & $2033.4^{+0.2}_{-0.3}$  \\
    \noalign{\smallskip}
    $a_c$ [au] & $\mathcal{U}(10.0,30.0)$  & $21.5^{+0.5}_{-0.4}$  \\
    \noalign{\smallskip}
    $e_c$  & $\mathcal{U}(0.0,1.0)$  & $0.114^{+0.002}_{-0.003}$  \\
    \noalign{\smallskip}
    $\omega_c$ [deg]  & $\mathcal{U}(0.0,360.0)$  & $133^{+2}_{-2}$  \\
    \noalign{\smallskip}
    $i_c$ [deg] & $\cos(i_d),\,\mathcal{U}(0.0,180.0)$ &  $40^{+2}_{-2}$ \\
    \noalign{\smallskip}
    $\Omega_c$ [deg] & $\mathcal{U}(0.0,360.0)$  & $105^{+3}_{-4}$  \\
    \noalign{\smallskip}
    $q_c$  & $\mathcal{U}(0.0,0.1)$  & $0.0179^{+0.0007}_{-0.0003}$  \\
    \noalign{\smallskip}
    $M_{\rm c}$ [M$_\mathrm{Jup}$] & (derived) & $19.4^{+0.8}_{-0.5}$  \\
    \noalign{\smallskip}
    \noalign{\smallskip}
    \hline
    \end{tabular}
\end{table}

\begin{figure}
    \centering
    \includegraphics[width=\linewidth]{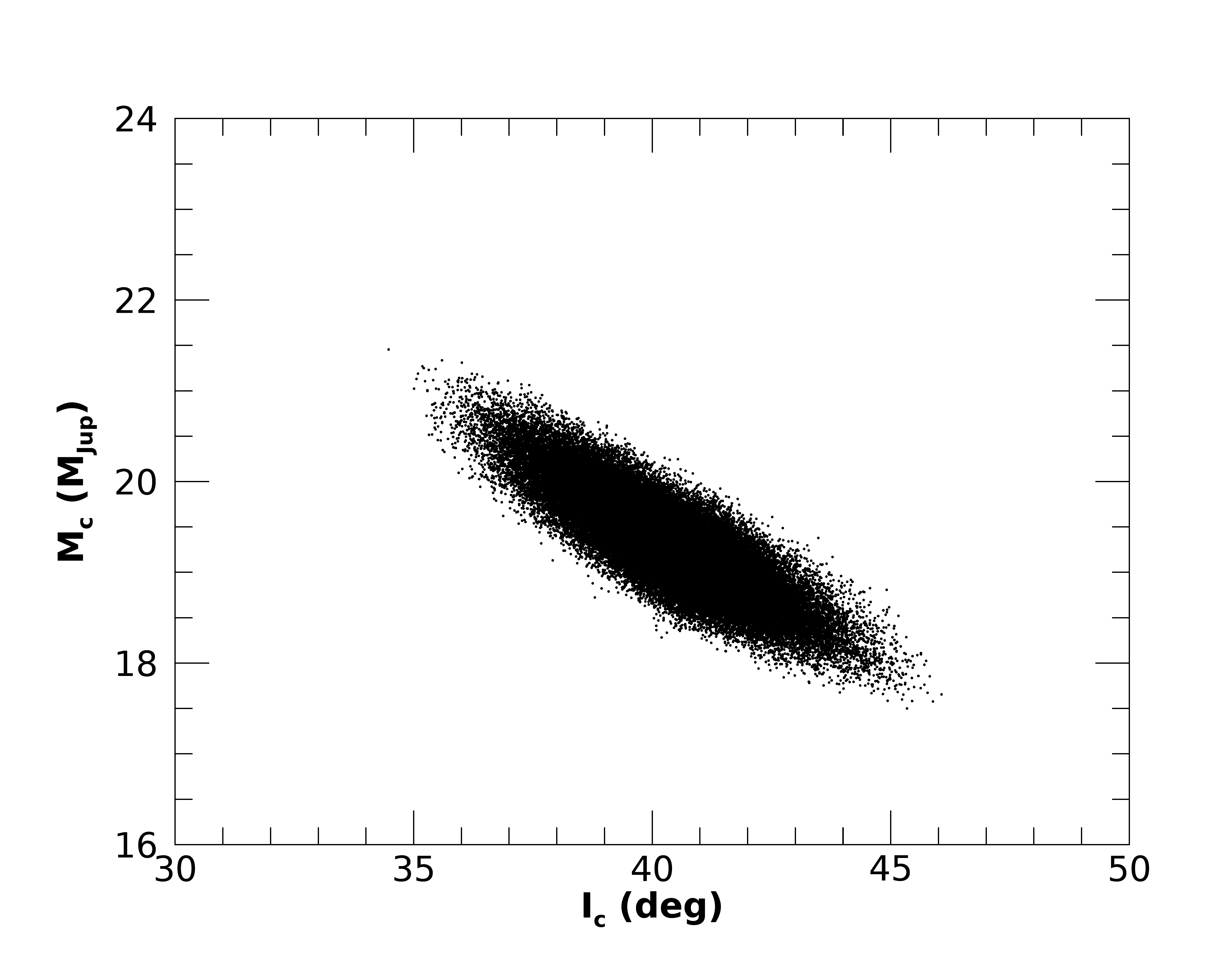} \\
    \includegraphics[width=\linewidth]{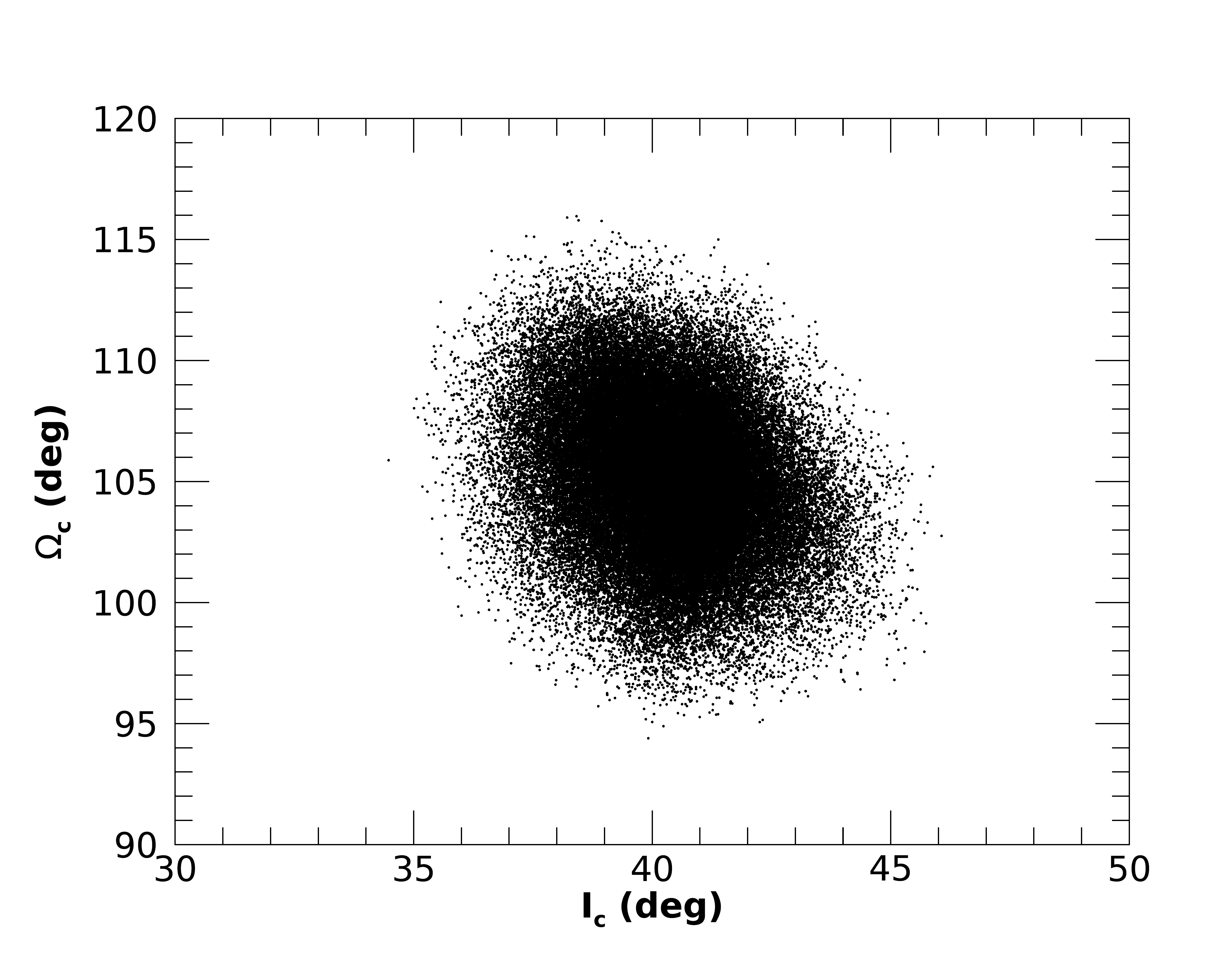}
    \caption{Joint posterior distributions for inclination, longitude of the
ascending node and true mass for HD\,72659 c.}
    \label{fig:jointdist_HD72659}
\end{figure}

Following the same approach adopted in the combined astrometry+RV analysis described in \cite{ruggieri2024b}, we ran the DE-MCMC analysis with the PMa+RV model fitted to the HIPPARCOS-Gaia absolute astrometry of HD\,72659 from Table \ref{tab:absastr3} and the offset-corrected RV dataset with the signal of HD\,72659 b removed. We do note that, with $a_b=4.60$ au, the orbital separation of HD\,72659 b, falls in the region of highest sensitivity of the PMa technique. However, based on the analytical formulation of \citet{Kervella2019}, at $a\simeq5$ au the measured PMa should be due to a companion with $M_{\rm b}=19\pm6$ \mjup\ (see also Fig. \ref{fig:hd72659_pma_imaging}). Given the $m_{\rm b}\sin(i)$ value obtained by the RV-only analysis, this would imply quite a low inclination orbit ($i\sim9$ deg). We therefore initially assumed that the observed PMa is entirely due to the effect of orbital motion induced by HD\,72659 c. The best-fit orbital solution for HD\,72659 c is reported in Table \ref{tab:fit_HD72659}. We obtain a formally precise solution, despite only seeing a long-term trend in the RV data covering $\sim20\%$ of the period. In particular, we obtained an orbital inclination $i_c = 40\pm2$ deg (for a retrograde orbit) and a mass ratio $q_c = 0.0179_{-0.0004}^{+0.0007}$, corresponding to a true mass of $M_{\rm c} = 19.4_{-0.5}^{+0.8}$ $M_\mathrm{Jup}$, formally identifying HD\,72659 c as a low-mass brown dwarf. The joint posteriors of the parameters mentioned above are shown in Fig. \ref{fig:jointdist_HD72659}. The true mass HD\,72659 c we obtain sits at the lower boundary of the 1-$\sigma$ mass interval predicted by the sensitivity curve in Fig. \ref{fig:hd72659_pma_imaging}, with a central true mass value at $a\sim20$ au closer to 30 $M_\mathrm{Jup}$. 

We also explored the possibility of a significant contribution to the observed PMa from HD\,72659 b by fitting a two-companion model to the combined RV+astrometry dataset. In particular, we ran the DE-MCMC analysis using the same broad priors for $i_b$ and $i_c$ and forcing $i_c$ in the range $10-15$ deg (or $165-180$ deg for a prograde orbit), the latter case corresponding to a range of true values of $M_{\rm b}$ that would encompass that indicated by the sensitivity curve shown in Fig. \ref{fig:hd72659_pma_imaging} given the semi-major axis of HD 72659 b's orbit. Even imposing the tight priors on the spectroscopically determined orbital elements for HD 72659 b, which are precisely determined by the RV-only analysis, convergence was not achieved in either case, with a vanishingly small value of $q_b$ and unconstrained values for $P_c$. We conclude that the signal of HD 72659 b is not present in the Hipparcos-Gaia absolute astrometry, therefore its true mass is likely within a factor of a few from its minimum value. In this respect, we cannot rule out a close to coplanar configuration between HD 72659 b and HD 72659 c, as in this case, assuming $i_b \sim40$ deg, the true mass of the former would be $M_{\rm b}\sim4.5$ \mjup, still far from being large enough to produce a detectable astrometric acceleration.

\subsection{Result from direct imaging}
\begin{figure*}
\centering
\includegraphics[width = \linewidth]{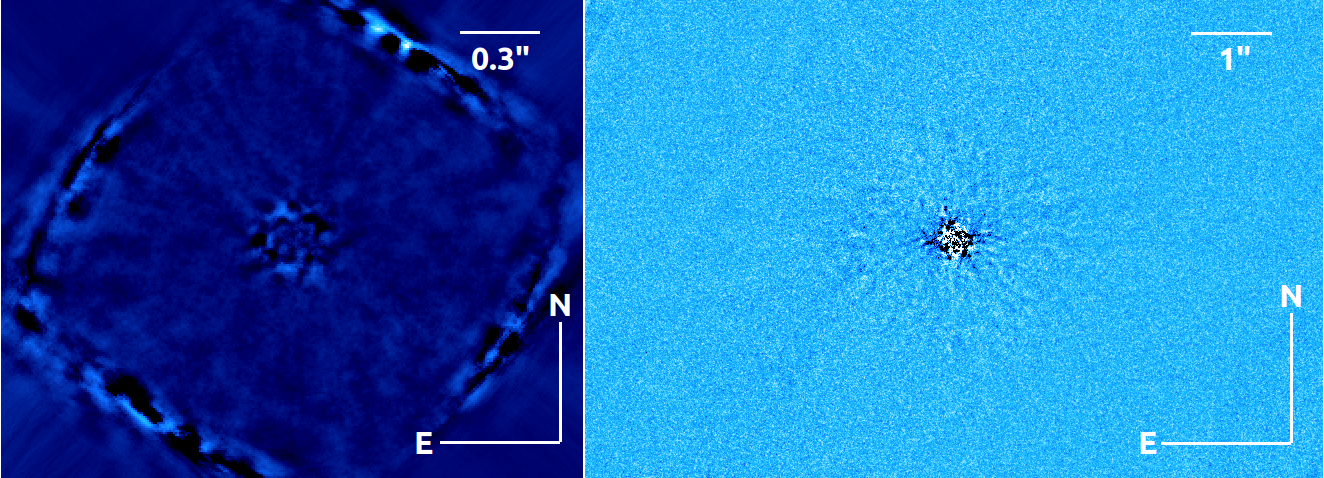}
\caption{Deepest imaging observations taken with SPHERE, both with IFS (left) and IRDIS (right).}
\label{fig:hd72659_irdis_ifs}
\end{figure*}

\begin{figure*}
%\vspace{-2cm}
   \centering
   \includegraphics[width = \linewidth]{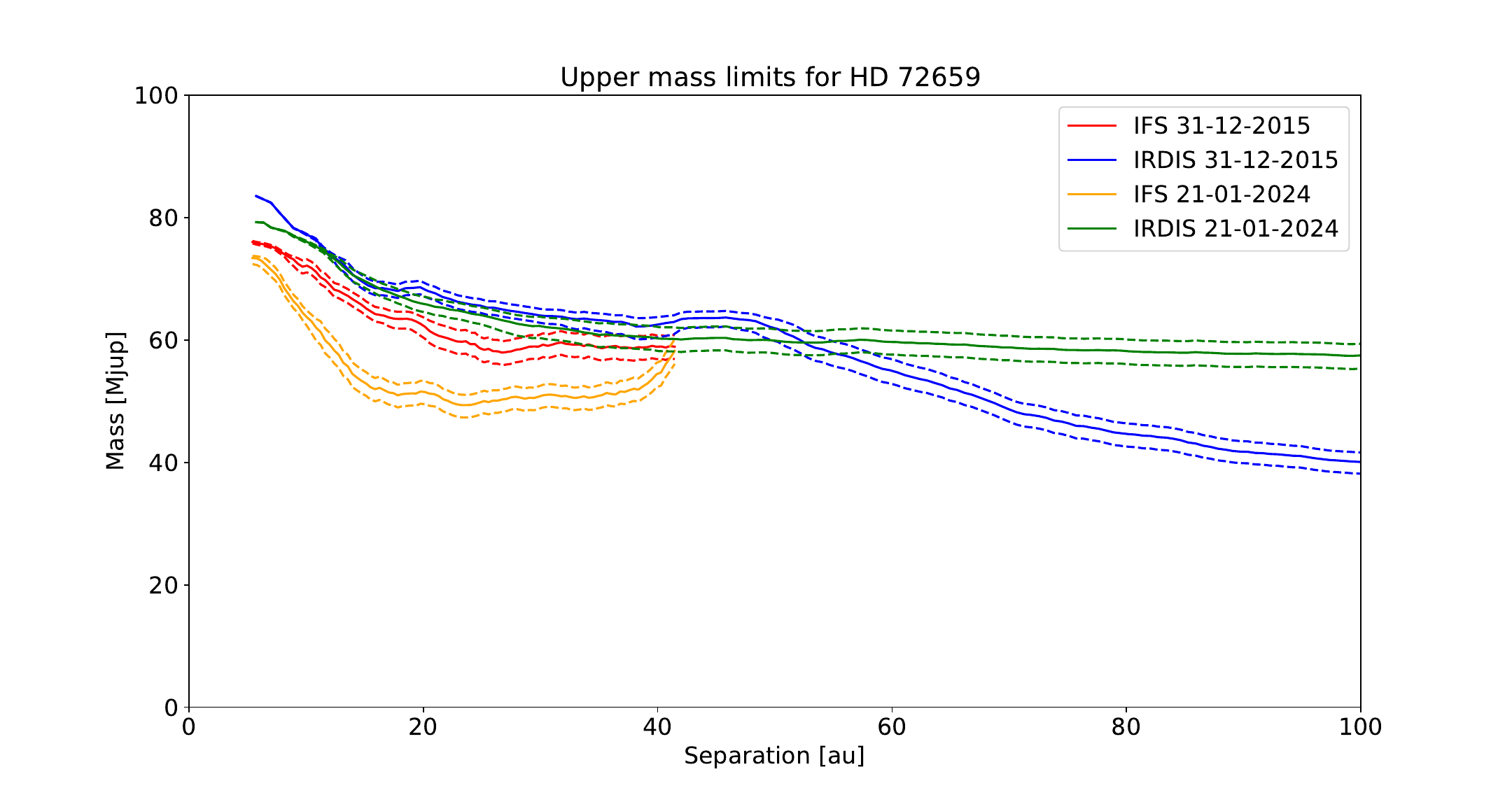}
      \caption{Upper mass limits for HD\,72659\,c, as a function of separation from the central star, derived from the previous and new SPHERE observations obtained using the AMES-COND models. The dashed lines represent the $1\sigma$ uncertainty.}
         \label{fig:compare_imaging}
   \end{figure*}

\begin{figure*}
%\vspace{-2cm}
   \centering
   \includegraphics[width = 0.9\linewidth]{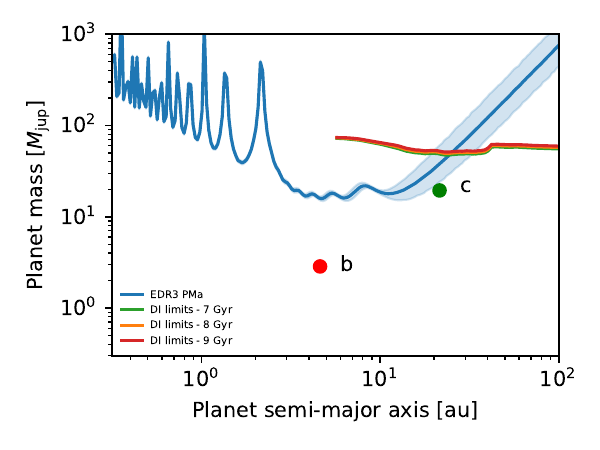}
      \caption{Upper limits in mass for the long-period substellar companion of HD\,72659 derived from our SPHERE observation and PMa sensitivity curves calculated following \citet{Kervella2022}. The positions of planet b (from RVs only) and planet c (as derived from our RV + astrometry analysis) are also shown.}
         \label{fig:hd72659_pma_imaging}
   \end{figure*}

We reduced our new SPHERE observation as described in Section \ref{sec:imagingdata}, the result is shown in Figure \ref{fig:hd72659_irdis_ifs}. As we can see, there is no clear trace of HD\,72659 c nor of any other stellar or sub-stellar companion.
We derived the contrast curve by calculating the standard deviation for pixels at the same separation from the central star and dividing this value by the normalization factor obtained using the offset PSF described in Section~\ref{sec:imagingdata}. We then evaluated the self-subtraction caused by the speckle subtraction method used by injecting in our dataset simulated planets with known contrast at different separations from the central star. These results were then used to correct the contrast previously obtained. Starting from the contrast we were then able to calculate the mass limits of possible companions around this star using the AMES-COND atmospheric models \citep{allard2003} and assuming an age of $8.0 \pm 1.0$ Gyr as derived in Section \ref{sec:specparam3}. Figure \ref{fig:compare_imaging} shows the results for both SPHERE observations using both IFS and IRDIS. As we can see, the new observation is deeper, especially with IFS. We note that the higher mass limits obtained at large separation with IRDIS in the new observation are not surprising, as they were obtained in K-band, where the background level is higher than in the H-band used for the previous observation. In this observation, we were more interested in searching companions at short separations from the host star, so the loss in sensitivity at larger separations was considered acceptable. Therefore, from now on, we will only consider the results obtained from our new observations, giving an upper limit on the mass of HD\,72659\,c of about 50 \mjup\ at the expected orbital separation for this object. Figure \ref{fig:hd72659_pma_imaging} contains the imaging detection limits, the PMa sensitivity curve as derived by \cite{Kervella2022}, and the position of the two planets. Since the object is not detected in SPHERE data, its position on this plot should be below (i.e., at lower masses than) the red line ($\sim 50$ \mjup). This is compatible with the astrometric result discussed in the previous section.

\subsection{The intriguing dynamics of the system}

Both planets b and c exhibit significant eccentricities, suggesting a complex evolutionary history. Mutual secular perturbations between the two planets in a coplanar configuration result in a maximum eccentricity of only 0.01 for each planet—far below the observed values. Two alternative explanations are possible: (1) the system may have experienced a phase of planet-planet scattering, during which one planet was ejected, leaving the remaining two on eccentric orbits; or (2) the two planets may have a significant mutual inclination, leading to Kozai–Lidov oscillations.
In the first scenario, following the planet–planet scattering event that increased the eccentricities to their current values, secular perturbations would cause modest eccentricity variations—ranging from 0.08 to 0.125 for planet c and from 0.22 to 0.29 for planet b, as resulting from a dedicated dynamical simulation in case of nearly complanar orbits. In contrast, the Kozai–Lidov mechanism would result in a much richer dynamical evolution. According to \cite{naoz2016}, this process can induce large variations in both eccentricity and inclination, with potential transitions between prograde and retrograde orbits, and even chaotic behavior.

\begin{figure}
    \centering
    \includegraphics[width=\linewidth]{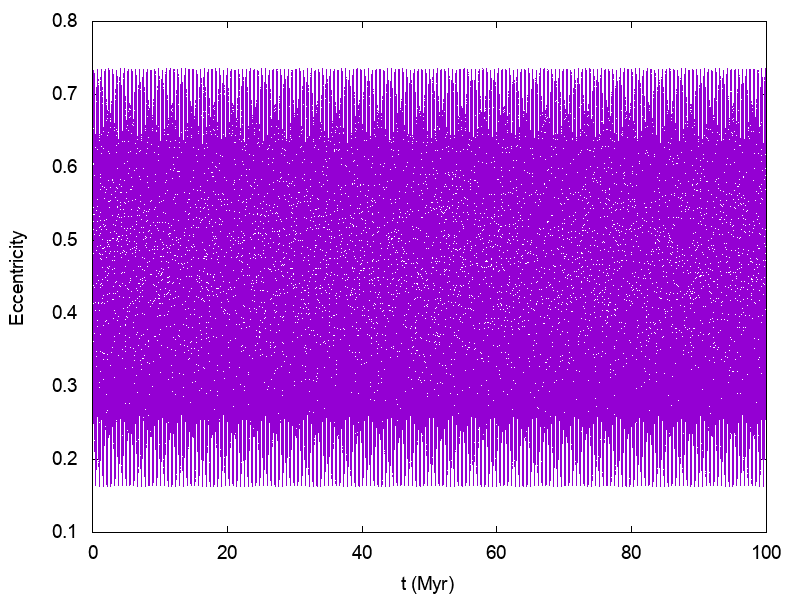} \\
    \includegraphics[width=\linewidth]{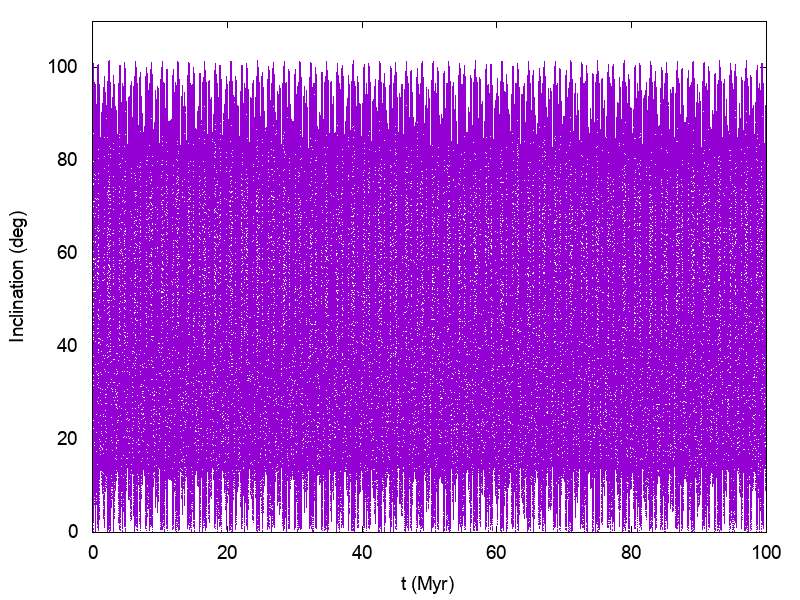}
    \caption{Evolution with time of the eccentricity (top panel) and inclination (bottom panel) of planet b perturbed by planet c.}
    \label{fig:dynamics}
\end{figure}

To explore the possible dynamics of the system under the assumption that it resides in a Kozai–Lidov state, we performed numerical integrations using the RADAU integrator \citep{radau1985}. The simulations employed the nominal masses and orbital parameters listed in Table \ref{tab:tabparam3} for planet b and Table \ref{tab:fit_HD72659} for planet c, adopting a mutual inclination of $50^\circ$. As illustrated in Fig. \ref{fig:dynamics}, the eccentricity of planet b varies between approximately 0.16 and 0.72, while its inclination oscillates between 0° and 100°, with the orbit periodically becoming retrograde. This behavior is consistent with the observed eccentricities of the planets.

Over the 100 Myr duration of the integration, the orbits exhibit semi-periodic evolution without signs of chaotic behavior. A similar dynamical pattern is observed when planet c is initially placed on a retrograde orbit. Random variations in the longitudes of pericenter and ascending node do not significantly affect the inclination evolution, although the eccentricity of planet b can be reduced, reaching a maximum of about 0.4 in some configurations.

The possible misalignment between the orbit of planet c and that of planet b may result from three distinct evolutionary scenarios. In the first, planet c formed concurrently with the primary star as a brown dwarf, while planet b formed later via core accretion within a circumstellar disk that was misaligned with planet c’s orbit \citep{batygin2012}.

A second viable scenario involves a massive circumstellar disk capable of producing several giant planets with varying masses. Subsequent planet–planet scattering among these bodies could have ejected some of the planets, leaving the survivors on highly inclined orbits \citep[][and references therein]{marzari2025}.

A third possibility is that the system experienced a stellar flyby within a dense stellar cluster during its early evolution. In this case, the inclination of planet c could have been excited by a strong gravitational interaction with a passing star \citep{mal2011,lad1998}. For this mechanism to be effective, the flyby must have occurred after the dissipation of the circumstellar disk and near the end of the cluster's lifetime; otherwise, the disk would have rapidly damped any inclination or eccentricity induced by the encounter \citep{marne2009}.

\section{Conclusion}
\label{sec:conclusions3}
In this work, we analyzed HD\,72659, putting severe constraints on the orbit of its two substellar companions. This was relatively easy for planet b, thanks to more than 20 years of high-precision RV data, but not as much for the recently discovered companion on the system's outer edges. Its period is so long that even with our large data set, we were not able to well constrain its orbit with RVs alone. However, the strong Hipparcos-Gaia PMa measured for this object greatly improved our results. In particular, we find a true mass that is compatible with \cite{feng2022} but, according to our results, they are underestimating the orbital period of HD\,72659 c by a factor of $\sim 2$ with a significance of $10.4\sigma$ (with respect to our RV+astrometry analysis). This is likely due to our larger data set that includes 91 data points gathered over 10 years. As we mentioned, the mass that we derive is lower than the value predicted by the Gaia sensitivity curve but formally compatible, and we excluded the possibility that planet b is also contributing to the astrometric signal. This would have implied an almost pole-on orbit, potentially casting doubts about the stability of a system with 2 very massive ($M \sim 20$ \mjup) objects. What's more, our new imaging data showed that there are no previously undetected stellar companions and confirmed that HD\,72659 c is, unfortunately, not detectable with SPHERE. Finally, we discussed the dynamics and possible origin of the system, which is of peculiar interest given that the inner planet has a moderate eccentricity while the outer companion is rather massive and not so distant. We concluded that the Kozai-Lidov mechanism might be at play in this system and that there is no chaotic evolution. This situation is somewhat similar to that of HD 11506 discussed in \cite{ruggieri2024b}. 

\begin{figure*}
    \centering
    \includegraphics[width=\textwidth]{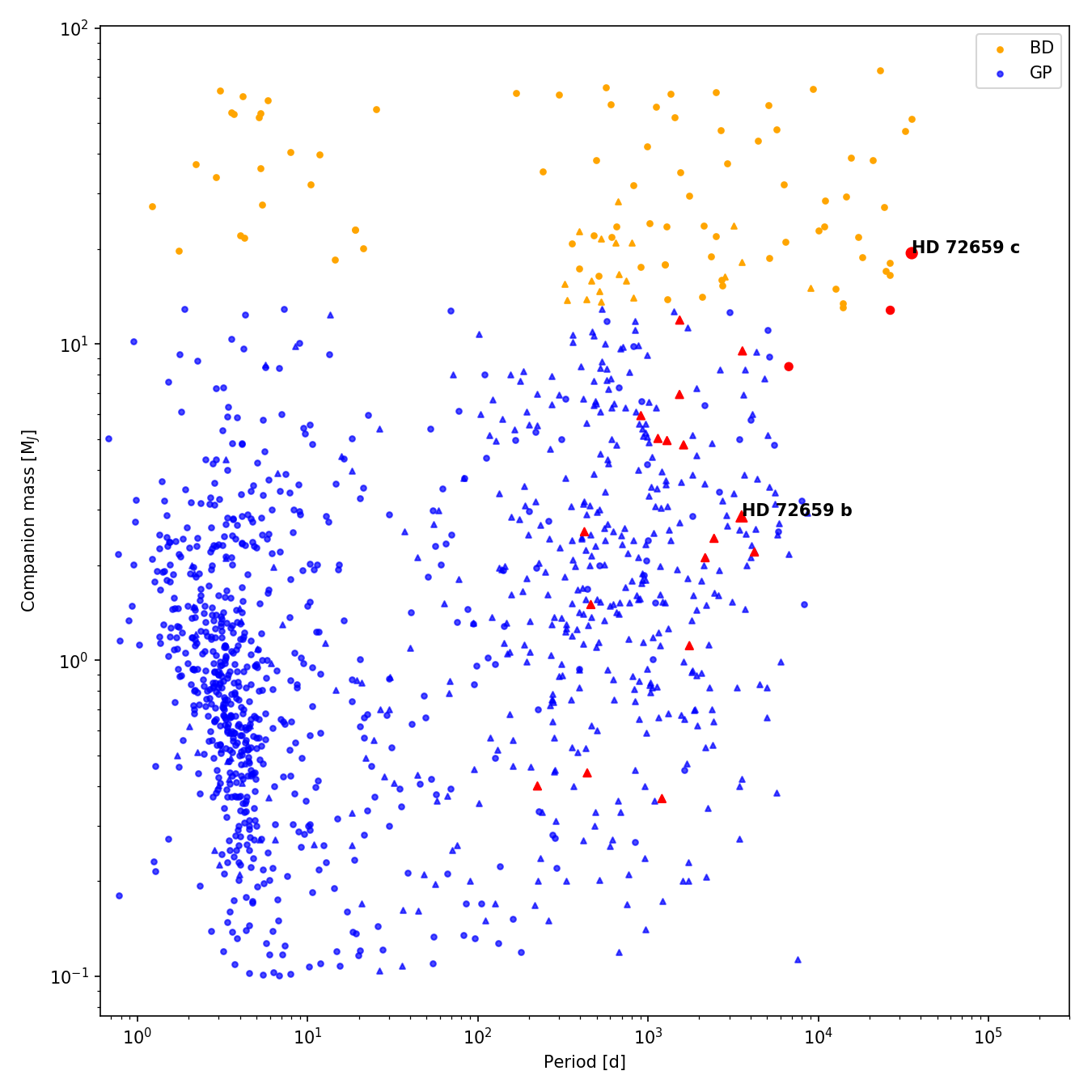}
    \caption{Mass-period plot of known exoplanets and BDs. Triangles indicate that only the minimum mass is known, while the true mass has been derived for objects indicated with dots. Blue indicates planets and orange indicates BDs (below or above 13 \mjup). All the GAPS KP objects are shown in red.}
    \label{fig:mass_period}
\end{figure*}

HD\,72659 c is placed in the so-called brown dwarfs desert, which is a region of the mass-period space where only a few objects are found. In particular, this system represents a precious case study because it also contains another planet with a high mass and a relatively long period. We note here that this star belongs to the KP sample selected by the GAPS team. This includes 16 stars for which an external gas giant was already known in 2012 and these were monitored with HARPS-N since then. Out of 16, three of them turned out to have another planet/BD at even longer periods (the other two are HD 75898 and HD 11506, see \citet{ruggieri2024b}). This is a significant number of outer companions and we will analyze the matter more in-depth in an upcoming dedicated statistical analysis (Pinamonti et al., in preparation). Figure \ref{fig:mass_period} shows the mass-period diagram of known exoplanets with our KP objects highlighted.

\begin{figure}
    \centering
    \includegraphics[width=\linewidth]{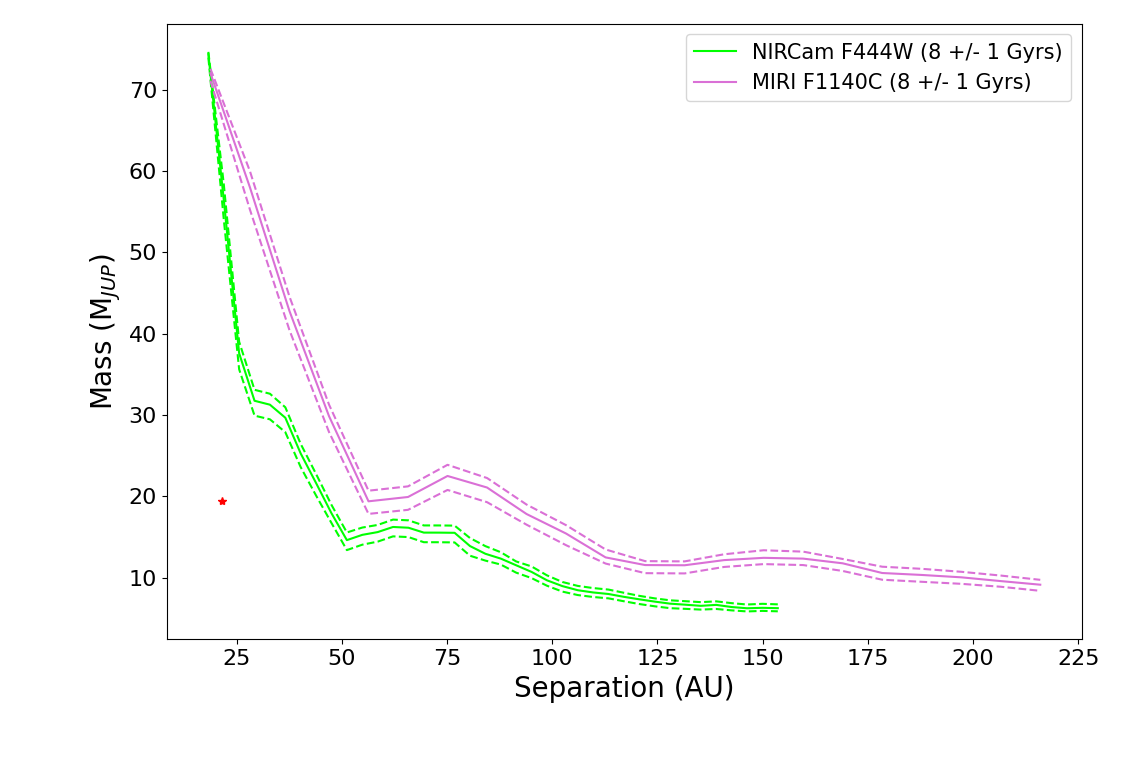}
    \caption{Theoretical detection limits for HD\,72659 using both the MIRI F1140C and the NIRCam F444W filters on JWST. The red star represents HD\,72659 c, while the dashed lines represent the $1\sigma$ uncertainties.}
    \label{fig:jwst}
\end{figure}

In the future, there is room for improvement in the analysis of this system.
Firstly, we will keep monitoring this target with HARPS-N, covering larger fractions of planet c's orbit. Secondly, Gaia DR4 will provide astrometric time series that will likely be sensitive to the mass and period range of HD\,72659 b, possibly helping to constrain the orbital inclination of this planet (and thus the true mass). As far as direct imaging is concerned, we investigated the potential of detecting the outer planet in this system with the James Webb Space Telescope (JWST) in the near- and mid-infrared. To this end, we computed theoretical contrast curves for MIRI F1140C filter ($\lambda =11.30$ $\mu$m, $\Delta \lambda = 0.8$  $\mu$m) and NIRCam F444W filter ($\lambda =4.402$ $\mu$m, $\Delta \lambda = 1.024$  $\mu$m) using  PanCAKE, a simulation tool that extends the official JWST exposure time calculator (Pandeia) to produce more accurate predictions of JWST coronagraphic performance \citep{Carter2}. We simulated a coronagraphic observation with both instruments, assuming an exposure time of 3600 seconds and a reference star of similar spectral type for Reference Differential Imaging \citep[RDI, ][]{Lafreniere} processing, which proved to be the most efficient post-processing technique for JWST coronagraphic observations \citep{Carter}. The contrast limits were then converted into mass limits using the atmospheric evolutionary model ATMO \citep{Phillips} and assuming a system age of $8 \pm 1$ Gyr. As shown in Fig. \ref{fig:jwst}, the outer planet lies below the detection thresholds of both instruments. We thus conclude that JWST cannot directly detect a planetary companion in such an old system. Nevertheless, future direct detection with the Extremely Large Telescope (ELT) could be possible. Combined with the well-determined metallicity and age of HD\,72659, the sum of these aspects in the coming years may allow us to shed further light on the architecture, dynamics, and composition of this system, making it a benchmark case study for directly imaged planetary systems. 

\section*{Acknowledgements}
This paper is supported by the Fondazione ICSC, Spoke 3 Astrophysics and Cosmos Observations. National Recovery and Resilience Plan (Piano Nazionale di Ripresa e Resilienza, PNRR) Project ID CN\_00000013 "Italian Research Center on    High-Performance Computing, Big Data and Quantum Computing"  funded by MUR Missione 4 Componente 2 Investimento 1.4: Potenziamento strutture di ricerca e creazione di "campioni nazionali di R\&S (M4C2-19 )" - Next Generation EU (NGEU).
We acknowledge the Italian center for Astronomical Archives (IA2, \url{https://www.ia2.inaf.it}), part of the Italian National Institute for Astrophysics (INAF), for providing technical assistance, services and supporting activities of the GAPS collaboration.
LM acknowledges financial contribution from PRIN MUR 2022 project
2022J4H55R.
MP acknowledges support from the European Union – NextGenerationEU (PRIN MUR 2022 20229R43BH) and the “Programma di Ricerca Fondamentale INAF 2023”.
TZi acknowledges support from CHEOPS ASI-INAF agreement n. 2019-29-HH.0, NVIDIA Academic Hardware Grant Program for the use of the Titan V GPU card and the Italian MUR Departments of Excellence grant 2023-2027 “Quantum Frontiers”.

%%%%%%%%%%%%%%%%%%%%%%%%%%%%%%%%%%%%%%%%%%%%%%%%%%
\section*{Data Availability}
The data underlying this article are available in the article and in its online supplementary material.

%%%%%%%%%%%%%%%%%%%% REFERENCES %%%%%%%%%%%%%%%%%%

% The best way to enter references is to use BibTeX:

\bibliographystyle{mnras}
\bibliography{bibliography} % if your bibtex file is called example.bib

% Alternatively you could enter them by hand, like this:
% This method is tedious and prone to error if you have lots of references
%\begin{thebibliography}{99}
%\bibitem[\protect\citeauthoryear{Author}{2012}]{Author2012}
%Author A.~N., 2013, Journal of Improbable Astronomy, 1, 1
%\bibitem[\protect\citeauthoryear{Others}{2013}]{Others2013}
%Others S., 2012, Journal of Interesting Stuff, 17, 198
%\end{thebibliography}

%%%%%%%%%%%%%%%%%%%%%%%%%%%%%%%%%%%%%%%%%%%%%%%%%%

%%%%%%%%%%%%%%%%% APPENDICES %%%%%%%%%%%%%%%%%%%%%
\newpage

\appendix
\section{HARPS-N time series of HD\,72659}
\label{sec:hd72659_timeseries}

%\setlength\tabcolsep{4 pt}
%\setlength\LTleft{-0.65cm}
%\centering
\vspace{-10cm}
\begin{table*}
\caption{\label{tab:hd72659_timeseries}RVs, S-index, and BIS time series for HD\,72659 derived from HARPS-N spectra, and their associated uncertainties.}
\begin{tabular}{ccccccc}
\hline
BJD - 24000000 &   RV [m/s] &   $\sigma_{\rm RV}$ [m/s] & \smw &   $\sigma_{\text{S}_{\text{MW}}}$ &  BIS [km/s] &  $\sigma_{\rm BIS}$ [km/s] \\
\hline
56295.7024 & -18128.52 & 0.36 & 0.152701 & 0.000281 & 0.01588 & 0.00050 \\
56296.6934 &  -18127.02 &           0.37 &   0.152241 &       0.000287 &      0.01597 &          0.00052 \\
56366.4057 &  -18130.07 &           0.89 &   0.155158 &       0.000972 &      0.01304 &          0.00126 \\
       56424.4005 &  -18129.20 &           0.62 &   0.154787 &       0.000586 &      0.02072 &          0.00087 \\
       56582.7663 &  -18134.66 &           0.49 &   0.157785 &       0.000414 &      0.02158 &          0.00069 \\
       56584.7678 &  -18134.47 &           0.65 &   0.157508 &       0.000635 &      0.03678 &          0.00092 \\
       56585.7432 &  -18136.68 &           0.57 &   0.158707 &       0.000526 &      0.02250 &          0.00080 \\
       56586.7532 &  -18136.37 &           0.54 &   0.158777 &       0.000484 &      0.02223 &          0.00076 \\
       56602.7091 &  -18134.41 &           0.57 &   0.159661 &       0.000514 &      0.01232 &          0.00080 \\
       56603.7258 &  -18136.61 &           0.44 &   0.158465 &       0.000362 &      0.02144 &          0.00062 \\
       56604.7585 &  -18137.50 &           0.51 &   0.157603 &       0.000460 &      0.02220 &          0.00073 \\
       56605.7157 &  -18138.02 &           0.90 &   0.157461 &       0.001026 &      0.02150 &          0.00127 \\
       56606.6921 &  -18137.43 &           0.83 &   0.159167 &       0.000922 &      0.01825 &          0.00117 \\
       56607.7349 &  -18136.55 &           0.60 &   0.158631 &       0.000572 &      0.01929 &          0.00085 \\
       56608.7073 &  -18136.42 &           0.53 &   0.159951 &       0.000471 &      0.02099 &          0.00075 \\
       56616.7400 &  -18137.23 &           0.66 &   0.159628 &       0.000648 &      0.02261 &          0.00094 \\
       56617.7609 &  -18134.72 &           0.45 &   0.159187 &       0.000382 &      0.02315 &          0.00064 \\
       56621.7206 &  -18138.08 &           0.64 &   0.158642 &       0.000628 &      0.02226 &          0.00090 \\
       56631.7167 &  -18138.12 &           0.49 &   0.157537 &       0.000420 &      0.01972 &          0.00069 \\
       56656.7345 &  -18139.59 &           1.11 &   0.155932 &       0.001460 &      0.01672 &          0.00157 \\
       56697.6168 &  -18141.18 &           0.42 &   0.157732 &       0.000334 &      0.02349 &          0.00059 \\
       56698.4638 &  -18143.82 &           0.68 &   0.158051 &       0.000660 &      0.02190 &          0.00096 \\
       56698.5242 &  -18142.04 &           0.64 &   0.159341 &       0.000602 &      0.01830 &          0.00090 \\
       56699.4345 &  -18145.23 &           0.46 &   0.156376 &       0.000403 &      0.02146 &          0.00066 \\
       56699.5379 &  -18143.74 &           0.51 &   0.156813 &       0.000462 &      0.02310 &          0.00073 \\
       56726.3793 &  -18143.02 &           1.40 &   0.153826 &       0.002053 &      0.02071 &          0.00199 \\
       56788.3844 &  -18147.78 &           0.62 &   0.155335 &       0.000651 &      0.02479 &          0.00088 \\
       56972.7223 &  -18157.58 &           0.54 &   0.157387 &       0.000530 &      0.02279 &          0.00076 \\
       56986.6989 &  -18158.13 &           0.52 &   0.158829 &       0.000505 &      0.02027 &          0.00073 \\
       56987.7020 &  -18153.68 &           0.81 &   0.159864 &       0.000918 &      0.02542 &          0.00114 \\
       56993.7418 &  -18155.02 &           0.97 &   0.162158 &       0.001221 &      0.02612 &          0.00137 \\
       56999.7099 &  -18157.21 &           0.69 &   0.160825 &       0.000767 &      0.02123 &          0.00098 \\
       57005.7391 &  -18161.98 &           1.23 &   0.160150 &       0.001801 &      0.01544 &          0.00173 \\
       57026.6625 &  -18155.51 &           1.52 &   0.165968 &       0.002547 &      0.02589 &          0.00215 \\
       57027.5805 &  -18155.50 &           0.67 &   0.159260 &       0.000701 &      0.01847 &          0.00095 \\
       57050.5464 &  -18161.18 &           0.94 &   0.157285 &       0.001187 &      0.01794 &          0.00133 \\
       57115.3487 &  -18162.54 &           0.39 &   0.157422 &       0.000336 &      0.01967 &          0.00055 \\
       57389.6412 &  -18180.94 &           0.60 &   0.154259 &       0.000610 &      0.02003 &          0.00084 \\
       57390.6323 &  -18179.71 &           0.59 &   0.154559 &       0.000570 &      0.01819 &          0.00083 \\
       57391.6100 &  -18182.02 &           0.56 &   0.154871 &       0.000544 &      0.01527 &          0.00080 \\
       57416.6833 &  -18180.06 &           0.72 &   0.155236 &       0.000802 &      0.01635 &          0.00101 \\
       57417.6616 &  -18177.79 &           1.80 &   0.150439 &       0.003411 &      0.02084 &          0.00254 \\
       57418.6265 &  -18182.60 &           1.36 &   0.153267 &       0.002173 &      0.02133 &          0.00193 \\
       57419.5859 &  -18179.18 &           1.56 &   0.159862 &       0.002586 &      0.01728 &          0.00220 \\
       57421.5918 &  -18183.27 &           1.23 &   0.153633 &       0.001688 &      0.02117 &          0.00174 \\
       57431.5913 &  -18181.88 &           0.72 &   0.156564 &       0.000767 &      0.01768 &          0.00101 \\
       57443.5071 &  -18180.74 &           0.84 &   0.156511 &       0.001005 &      0.02064 &          0.00119 \\
       57444.4874 &  -18185.17 &           0.67 &   0.157834 &       0.000701 &      0.01802 &          0.00095 \\
       57445.4249 &  -18182.52 &           0.69 &   0.157068 &       0.000748 &      0.01911 &          0.00098 \\
       57472.3970 &  -18182.10 &           1.16 &   0.155105 &       0.001665 &      0.01808 &          0.00164 \\
       57474.3927 &  -18179.85 &           0.53 &   0.156853 &       0.000503 &      0.02071 &          0.00075 \\
       57475.3540 &  -18183.60 &           0.69 &   0.158722 &       0.000726 &      0.02017 &          0.00097 \\
       57501.4038 &  -18183.15 &           0.54 &   0.156733 &       0.000518 &      0.02072 &          0.00076 \\
       57506.4092 &  -18188.49 &           1.21 &   0.151329 &       0.001731 &      0.02170 &          0.00171 \\
       57525.3663 &  -18188.15 &           0.89 &   0.154391 &       0.001100 &      0.01650 &          0.00126 \\
       57526.3710 &  -18185.35 &           1.04 &   0.153510 &       0.001432 &      0.01872 &          0.00147 \\
       57679.7588 &  -18195.32 &           0.62 &   0.153308 &       0.000631 &      0.02219 &          0.00087 \\
       57680.7602 &  -18197.46 &           0.59 &   0.153744 &       0.000589 &      0.01832 &          0.00084 \\
       57681.7589 &  -18194.74 &           0.41 &   0.154710 &       0.000351 &      0.01870 &          0.00058 \\
       57756.6337 &  -18202.42 &           1.12 &   0.154860 &       0.001505 &      0.01646 &          0.00159 \\
       57757.6547 &  -18197.90 &           0.56 &   0.155444 &       0.000523 &      0.01775 &          0.00079 \\
       57770.6198 &  -18196.68 &           0.52 &   0.156766 &       0.000474 &      0.02266 &          0.00073 \\
\hline
\end{tabular}
\end{table*}

\begin{table*}
\contcaption{}
\label{tab:continued}
\begin{tabular}{ccccccc}
\hline
BJD - 24000000 &   RV [m/s] &   $\sigma_{\rm RV}$ [m/s] & \smw &   $\sigma_{\text{S}_{\text{MW}}}$ &  BIS [km/s] &  $\sigma_{\rm BIS}$ [km/s] \\
\hline
       57771.5350 &  -18200.09 &           0.53 &   0.157223 &       0.000503 &      0.01864 &          0.00076 \\
       57772.5843 &  -18202.04 &           0.53 &   0.157463 &       0.000483 &      0.01947 &          0.00075 \\
       57782.5427 &  -18198.74 &           0.54 &   0.156459 &       0.000498 &      0.01881 &          0.00076 \\
       57809.5436 &  -18199.28 &           0.80 &   0.155480 &       0.000882 &      0.01916 &          0.00113 \\
       57810.4063 &  -18198.77 &           1.90 &   0.158130 &       0.003476 &      0.02177 &          0.00268 \\
       57881.3709 &  -18207.43 &           0.41 &   0.154218 &       0.000352 &      0.01818 &          0.00059 \\
       58142.6535 &  -18213.58 &           1.45 &   0.147257 &       0.006138 &      0.01983 &          0.00205 \\
       58190.4917 &  -18218.05 &           0.50 &   0.155321 &       0.000434 &      0.01715 &          0.00071 \\
       58478.7042 &  -18214.26 &           0.57 &   0.159368 &       0.000541 &      0.01633 &          0.00081 \\
       58495.7154 &  -18213.67 &           0.56 &   0.160771 &       0.000515 &      0.02045 &          0.00079 \\
       58503.6342 &  -18212.19 &           0.74 &   0.160655 &       0.000770 &      0.01653 &          0.00104 \\
       58506.5572 &  -18211.81 &           0.91 &   0.161582 &       0.001037 &      0.02090 &          0.00128 \\
       58855.6493 &  -18188.92 &           1.26 &   0.151742 &       0.001627 &      0.01968 &          0.00179 \\
       59516.7649 &  -18163.98 &           0.54 &   0.152952 &       0.000540 &      0.01507 &          0.00077 \\
       59566.5909 &  -18161.92 &           0.49 &   0.158129 &       0.000485 &      0.01837 &          0.00069 \\
       59585.6139 &  -18165.63 &           1.02 &   0.157337 &       0.001487 &      0.02041 &          0.00144 \\
       59620.6060 &  -18160.70 &           1.24 &   0.162232 &       0.002115 &      0.01077 &          0.00175 \\
       59644.5626 &  -18159.35 &           0.58 &   0.163620 &       0.000622 &      0.02150 &          0.00082 \\
       59648.4823 &  -18161.16 &           0.48 &   0.162087 &       0.000468 &      0.02214 &          0.00068 \\
       59683.4372 &  -18159.54 &           1.07 &   0.161488 &       0.001607 &      0.01800 &          0.00152 \\
       59686.4390 &  -18164.63 &           0.55 &   0.161180 &       0.000569 &      0.02089 &          0.00078 \\
       59714.3661 &  -18162.75 &           0.41 &   0.158214 &       0.000370 &      0.01961 &          0.00058 \\
       59867.7434 &  -18171.86 &           0.47 &   0.156610 &       0.000433 &      0.01672 &          0.00067 \\
       59871.7500 &  -18172.58 &           0.82 &   0.155413 &       0.000981 &      0.02142 &          0.00116 \\
       59895.6624 &  -18172.48 &           0.52 &   0.153535 &       0.000501 &      0.01695 &          0.00073 \\
       59995.6094 &  -18175.35 &           0.68 &   0.160086 &       0.000753 &      0.01809 &          0.00096 \\
       60013.5062 &  -18180.03 &           0.75 &   0.154697 &       0.000886 &      0.01665 &          0.00106 \\
       60018.5418 &  -18178.63 &           0.51 &   0.156177 &       0.000500 &      0.01489 &          0.00072 \\
       60046.4007 &  -18177.06 &           0.86 &   0.159575 &       0.001081 &      0.01715 &          0.00121 \\
\hline
\end{tabular}
\end{table*}

%%%%%%%%%%%%%%%%%%%%%%%%%%%%%%%%%%%%%%%%%%%%%%%%%%

% Don't change these lines
\bsp	% typesetting comment
\label{lastpage}
\end{document}